\documentclass{aa}  

\usepackage{graphicx}
\usepackage[dvipsnames]{xcolor}
\usepackage{multirow}

\usepackage{txfonts}
\usepackage[]{hyperref}
\hypersetup{
    colorlinks=true,
    citecolor=blue,
    linkcolor=blue,
    urlcolor=blue,
    }

\newcommand{\orcid}[1]{\href{https://orcid.org/#1}{\includegraphics[width=10pt]{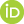}}}

\defcitealias{ferrari2024}{F24}

\begin{document}

   \title{SN 2023ixf: interaction signatures in the spectrum at 445 days}

   \author{Gastón Folatelli\inst{1,2,3}\orcid{0000-0001-5247-1486}
          \thanks{gaston@fcaglp.unlp.edu.ar}
          \and
          Lucía Ferrari\inst{1,2}\orcid{0009-0000-6303-4169}
          \and
          Keila Ertini\inst{1,2}\orcid{0000-0001-7251-8368}
          \and
          Hanindyo Kuncarayakti\inst{4,5}\orcid{0000-0002-1132-1366}
          \and
          Keiichi Maeda\inst{6}\orcid{0000-0003-2611-7269}
          }

   \institute{Instituto de Astrofísica de La Plata (IALP), CONICET -- UNLP, La Plata, Argentina
         \and
             Facultad de Ciencias Astronómicas y Geofísicas, Universidad Nacional de La Plata (UNLP), Av.\ Centenario s/n B1900FWA, La Plata, Argentina
        \and
            Kavli Institute for the Physics and Mathematics of the Universe (WPI), The University of Tokyo Institutes for Advanced Study, The University of Tokyo, Kashiwa, 277-8583 Chiba, Japan
        \and
            Tuorla Observatory, Department of Physics and Astronomy, 20014 University of Turku, Finland
        \and
            Finnish Centre for Astronomy with ESO (FINCA), 20014 University of Turku, Finland
        \and
            Department of Astronomy, Kyoto University, Kitashirakawa-Oiwake-cho, Sakyo-ku, Kyoto 606-8502, Japan
             }

   \date{Received Month 02, 2025; accepted Month 05, 2025}

  \abstract
   {SN~2023ixf is one of the nearest and brightest Type II supernovae (SNe) of the past decades. Its proximity and extremely early discovery has allowed a great number of studies based on extensive observations throughout the electromagnetic spectrum. A rich set of pre-explosion data provided important insight on the properties of the progenitor star. There has been, however, a wide range of estimated initial masses of $9-22$~M$_\odot$. Early monitoring of the SN also showed the presence of a dense CSM structure near the star ($\lesssim$$10^{15}$~cm) that was probably expelled in the last years prior to the explosion. At further distances, there have been indications of a drop in the CSM density. These extended CSM structure can be further probed with late-time observations during the nebular phase.}
   {We monitor the spectroscopic evolution of SN~2023ixf at late phases with the aim of characterizing the progenitor properties. The observations also serve to search for indications of ejecta--CSM interaction that may shed light on the mass-loss processes during the final stages of evolution of the progenitor star.}
   {This study is based on a nebular spectrum obtained with GMOS at the Gemini North Telescope 445 days after explosion. The SN evolution is analyzed in comparison with a previous spectrum at an age of 259 days. The 445-d spectrum is further compared with those of similar SNe~II and with synthetic radiation-transfer nebular spectra. Line profiles are used to determine properties of the emitting regions. [\ion{O}{i}] and [\ion{Ca}{ii}] line fluxes are used to derive an estimate of the progenitor mass at birth.}
   {The 445-d spectrum exhibits a dramatic evolution with { signs} of ejecta--CSM interaction. The H$\alpha$ profile shows a complex profile that can be separated into a boxy component {possibly arising} from the interaction with a CSM shell and a central peaked component that may be due to the radioactive-powered SN ejecta. The CSM shell would be located at a distance of $\sim$$10^{16}$~cm from the progenitor and it may be associated with mass loss occurring up until $\approx$$500-1000$ years before the explosion. Similar interaction signatures have been detected in other SNe~II, although for events with standard plateau durations this happened at times later than $600-700$ days. SN~2023ixf appears to belong to a group of SNe~II with short plateaus or linear light curves that develop interaction features before $\approx$500 days. Other lines, such as those from [\ion{O}{i}] and [\ion{Ca}{ii}] appear to be unaffected by the CSM interaction. This allowed us to estimate an initial progenitor mass, which resulted in the relatively low range of $10-15$~M$_\odot$.}
   {}

   \keywords{supernovae:general --- supernovae: individual: SN~2023ixf}

   \maketitle

\section{Introduction}\label{sec:intro}

Mass loss plays a critical role in the evolution and final fate of massive stars. The structure of the star at core collapse as well as the distribution of circumstellar material (CSM) due to mass-loss processes dictates the observational properties of the ensuing supernova (SN) event. This means that studying SNe can provide important insight on this issue. 

The most common type of core-collapse SN is that with hydrogen-dominated spectra, known as Type II SNe (SNe~II). These events have been associated with red supergiant (RSG) stars that retain most of their H-rich envelopes \citep{Smartt2009,vandyk2012,smartt2015}. Historically it appeared that SN~II progenitors suffered only mild mass loss during their evolution. However, in the past decade there has been amounting evidence of substantial mass loss at least in the final stages of evolution \citep[e.g. see][]{Forster2018}. 

There is a variety of interaction signatures that can be identified through early observations of SNe~II. It can be high luminosity, especially in the ultraviolet, and blue colors \citep{Dessart2022,Jacobson-Galan2024,Moriya2023}, a delayed rise to maximum light \citep{Moriya2018,Forster2018}, radio and X-ray emission from non-thermal processes in the interaction region \citep{Chevalier2017,Matsuoka2019,Moriya2021}, high-velocity absorption features in the spectra \citep{Chugai2007,Gutierrez2017,Dessart2022}, and short-duration narrow emission lines in the optical spectra during the first days/week after the explosion \citep[also known as "flash features"][]{Yaron2017,Bruch2023,Jacobson-Galan2024}. 

These observations probe CSM structures that typically extend to $10^{14}-10^{15}$~cm and must be formed during the final months to years prior to the explosion, given the typical RSG wind velocities. 
The associated mass-loss rates are $\dot{\mathrm{M}}\sim\,10^{-3}-10^{-1}$~M$_\odot$~yr$^{-1}$, which is orders of magnitude larger than what is measured in RSG stars \citep{Beasor2020} and than what is predicted by standard wind-powering mechanisms in isolated stars \citep{Vink2001}. This discrepancy has led to suggest alternative mass-loss mechanisms via accelerated winds \citep{Moriya2017}, episodic outbursts or mass transfer in interacting binaries \citep[see a review in][]{Smith2014}. 

Late-time indications of ejecta--CSM interaction have also been reported for SNe~II in the form of complex, flat-topped or boxy\footnote{{That is, usually broad emission exhibiting sharp, roughly symmetrical edges.}} H$\alpha$ emission profiles, usually accompanied by a flattening of the light curves due to the extra power released by the interaction. There is a wide range of ages when this interaction signatures are detected. For several SNe~II with particularly short plateaus or linear/fast-declining light curves, this behavior was observed to start as soon as $\sim$$200-500$ days after the explosion {\citep[several examples are listed in Section~\ref{sec:inter} and in the Introduction sections of][]{Weil2020,Dessart2023}}. This is in contrast with the lack of boxy or flat-topped H$\alpha$ profiles reported at ages between $\approx$150 and 500 days among 38 normal-plateau SNe~II in the sample of \citet{Silverman2017}\footnote{Where the authors specifically excluded linear SNe~II.}. Only a handful of SNe~II with standard plateau durations have shown similar signatures but at ages later than $\approx$600 days. Given the dimness of SNe at ages of several hundred days and later, and the ensuing lack of coverage, it is currently not possible to establish the actual fraction of SNe~II that undergo this transformation from radioactive-powered to interaction-powered evolution. However, because mass loss is ubiquitous among massive stars, this transition should eventually occur in all core-collapse SNe \citep{Dessart2022}. Based on this, the question should be about the timescale of the transition rather than about the fraction of objects that suffer it. 

The late-time interaction features probe CSM structures located at further distance of $\sim$$10^{15}-10^{16}$~cm from the progenitor star.
This material must have been expelled by the star during the final centuries before the explosion. The appearance of box-like H$\alpha$ profiles in the standard-plateau SNe~II SN~2004et and SN~2017eaw has been associated with mass loss rates of $\dot{\mathrm{M}}\sim10^{-6}$~M$_\odot$~yr$^{-1}$, in agreement with estimates from radio and X-ray observations \citep{Kotak2009,Weil2020}. This is well in line with the estimated mass loss rates for RSG stars.

By introducing a constant power source due to interaction in radiative transfer calculations of SNe~II, \citet{Dessart2022} were able to qualitatively reproduce the photometric and spectroscopic features attributed to interaction, such as enhanced UV emission, high-velocity absorptions during the plateau phase, and the development of boxy H$\alpha$ profiles during the nebular phase. \citet{Dessart2023} extend the analysis to later times and find a good agreement with the overall evolution of several SNe~II that showed signs of interaction in the form of boxy H$\alpha$ emission and light-curve flattening.

In the present work we analyze a nebular spectrum of the nearby Type II SN~2023ixf obtained at 445 days after the explosion. The new spectrum shows a dramatic evolution in comparison with the 259-d nebular spectrum published by \citet{ferrari2024} \citepalias[][hereafter]{ferrari2024}, especially in the appearance of a boxy H$\alpha$ profile. While the present analysis was under development, an additional spectrum of SN~2023ixf at 363 days was made public by \citet{Kumar2025} where the boxy H$\alpha$ profile had already emerged. {More recently, \citet{Michel2025}, \citet{Zheng2025arXiv250313974Z}, and \citet{Li2025arXiv250403856L} presented thorough spectroscopic follow-up observations reaching until 413, 442, and 407 days, respectively, where this evolution of the H$\alpha$ profile is observed.}
This clearly indicates that SN~2023ixf belongs to the group of SNe~II that showed interaction signatures during the nebular phase prior to $\sim$500 days. This is in line with the relatively short plateau phase of $\approx$80 days that was observed for this SN \citep{bersten2024}. 

Given its proximity, SN~2023ixf has provided a wealth of observational data that shed light on the nature of its progenitor star and properties of its CSM. Early-time observations indicated the presence of a dense CSM confined within $\lesssim$$10^{15}$~cm that was associated with enhanced mass loss with rates of $\dot{\mathrm{M}}\sim10^{-3}-10^{-2}$~M$_\odot$~yr$^{-1}$ during the last years prior to the explosion {\citep{teja2023,jacobsongalan2023,bostroem2023,Zhang2023,Li2024,Zimmerman2024}}.
Similar mass-loss rates were favored by hydrodynamical modeling of the early-time light curves \citep{Martinez2024,Zimmerman2024,Moriya2024}, although higher values of up to 1~M$_\odot$~yr$^{-1}$ were also suggested \citep{hiramatsu2023}. In contrast, lower mass-loss rates of $\dot{\mathrm{M}}\sim10^{-4}$~M$_\odot$~yr$^{-1}$ were indicated based on early X-ray luminosities \citep{grefenstette2023,Chandra2024}, from an analysis of the progenitor candidate spectral energy distribution \citep[SED;][]{Qin2024}, {and from the H$\alpha$ recombination luminosity \citep{Zhang2023}}. However, these lower values are in tension with limits placed by non-detections in the millimeter wavelength that exclude the range between $\sim$$10^{-6}$ and $\sim$$10^{-2}$~M$_\odot$~yr$^{-1}$ \citep{berger2023}.

At further distances from the progenitor ($\gtrsim$$10^{15}$~cm) the CSM density appears to drop, as evidenced by the disappearance of the narrow emission lines followed by an increase in the X-ray flux after day 4 \citep{Zimmerman2024}, by the continued drop in ultraviolet emission \citep{Bostroem2024}, and by late-time radio detections \citep{Iwata2025}. This may indicate that a phase of lower mass-loss rate occurred during the decades prior to explosion, as also suggested by the lack of outbursts shown by the progenitor candidate in pre-explosion images \citep{jencson2023,Dong2023,neustadt2024,Ransome2024,vandyk2024}. The question remains about how the CSM properties extend to even further distances ($\gtrsim$$10^{16}$~cm) that are associated with the mass-loss processes during the final centuries or millennia of the progenitor star. 

Regarding the progenitor star of SN~2023ixf, there has been numerous studies on the progenitor candidate detection in pre-explosion observations. The concensus is that it was a dust-enshrouded, variable RSG star. However, there has been a large variety of derived zero-age main sequence masses spanning the range of M$_\mathrm{ZAMS}\approx9-20$~M$_\odot$\footnote{Excluding the estimate of M$_\mathrm{ZAMS}=8-10$~M$_\odot$ by \citet{pledger2023} that neglects the presence of dust.} 
(\citealt[][$11\pm2$~M$_\odot$]{kilpatrick2023}; \citealt[][$17\pm4$~M$_\odot$]{jencson2023}; \citealt[][$16.2-17.4$~M$_\odot$]{niu2023}; \citealt[][$9-14$~M$_\odot$]{neustadt2024}; \citealt[][$12^{+2}_{-1}$~M$_\odot$]{Xiang2024}; \citealt[][$14-20$~M$_\odot$]{Ransome2024}; \citealt[][$12-15$~M$_\odot$]{vandyk2024}; {\citealt[][$18.2^{+1.3}_{-0.6}$~M$_\odot$]{Qin2024}}). Comparatively higher initial masses of the progenitor were derived from the analysis of the surrounding stellar populations by \citealt[][($17-19$~M$_\odot$)]{niu2023}, and \citealt[][($\approx$$22$~M$_\odot$)]{liu2023}. A study of the periodic variability observed for the progenitor candidate by \citet{soraisam2023} also favored a high initial mass of M$_\mathrm{ZAMS}=20\pm4$~M$_\odot$. Hydrodynamical modeling of the SN light curves by \citet{bersten2024} provided support to a relatively low-mass progenitor with M$_\mathrm{ZAMS}<15$~M$_\odot$ \citep[see also][]{Moriya2024}. Alternative hydrodynamical models by \citet{Fang2025} based on progenitors with enhanced mass loss suggested a slightly higher initial mass of M$_\mathrm{ZAMS}=15-16$~M$_\odot$.

Nebular spectroscopy of SNe~II can provide important independent information about the initial mass of the progenitor. Synthetic spectra show a dependence of the [\ion{O}{i}]\,$\lambda\lambda\,6300,6364$ luminosity with progenitor mass \citep{Jerkstrand2012,Dessart2021}. The flux ratio of [\ion{O}{I}]\,$\lambda\lambda\,6300,6364$ over [\ion{Ca}{ii}]\,$\lambda\lambda\,7291,7324$ has been proposed as an indicator of the CO core mass and thereby of M$_\mathrm{ZAMS}$ \citep{Fransson1987,Fransson1989}. Based on this, \citetalias{ferrari2024} derived a progenitor mass of M$_\mathrm{ZAMS}=12-15$~M$_\odot$ from the 259-d spectrum of SN~2023ixf. 

Alongside the study of the CSM interaction signatures, here we use the 445-d spectrum of SN~2023ixf to revisit the question of the progenitor mass. In Section~\ref{sec:obs} we present the observations used in this work. Section~\ref{sec:late-ev} provides an analysis of the data with focus on the interaction signatures. Section~\ref{sec:prog_mass} presents an updated estimate of the progenitor mass. Finally, our conclusions are given in Section~\ref{sec:conclusion}. 

\section{Observational data} \label{sec:obs}

\begin{figure*}
    \centering
    \includegraphics[width=\linewidth]{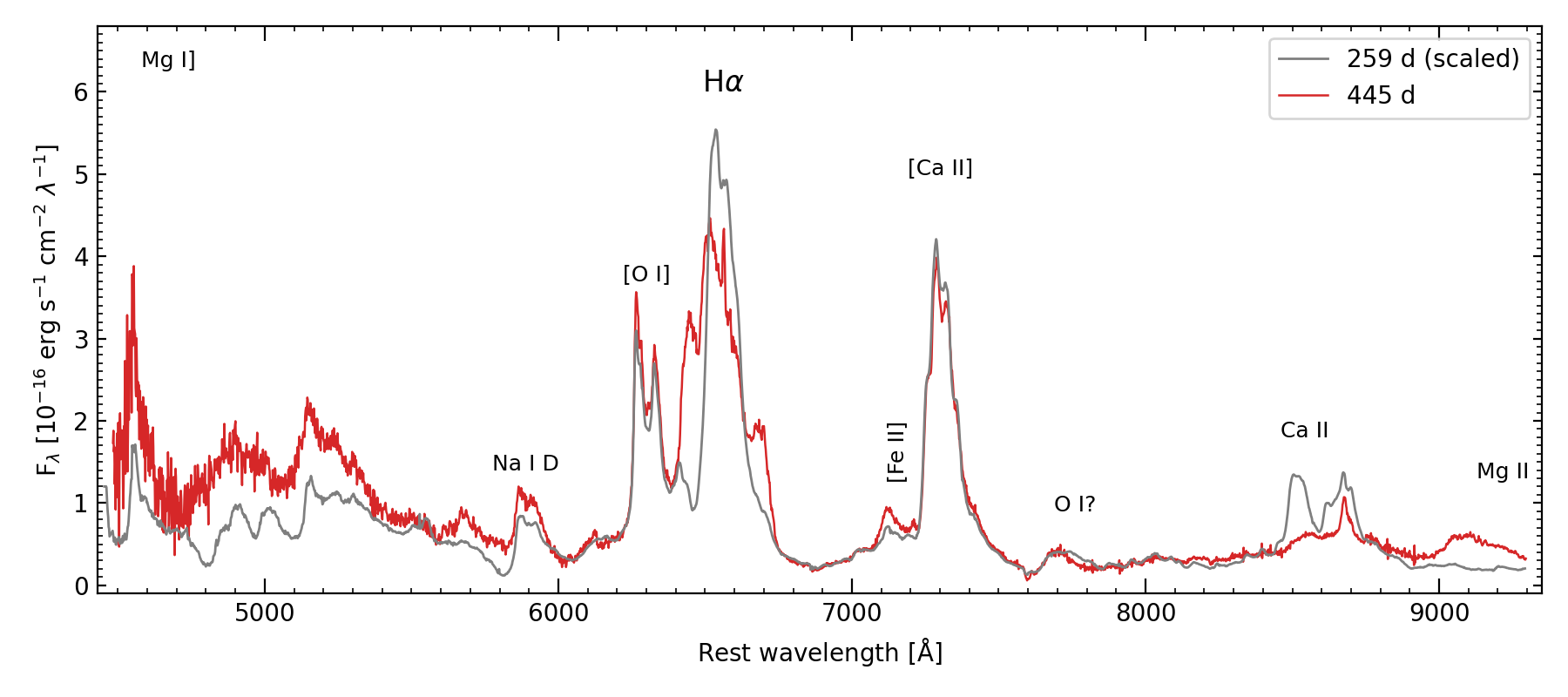}
    \caption{Nebular spectrum of SN~2023ixf at 445 days (red line) compared with the spectrum at 259 days \citepalias[gray line;][]{ferrari2024}. The spectra were corrected by extinction and multiplied by respective constant factors to make them match the $R$-band magnitude at 445 days (see text). The main spectral features are identified with the corresponding ion.}
    \label{fig:nebspec}
\end{figure*}

The present analysis is based on a nebular spectrum of SN~2023ixf obtained with the Gemini Multi-Object Spectrograph \citep[GMOS;][]{hook2004} mounted on the Gemini North Telescope (program GN-2024A-Q-309; PI Ferrari) on 2024 August $5.28$ UT. The observations were made in long-slit mode using the R400 grating, which provided a spectral resolution of {$\approx$600 as measured on several sky lines}. This is the same instrumentation and setup as that employed in previous observations of the same SN by \citetalias{ferrari2024}. The new observations were divided into six 700-s exposures. The data were processed through standard procedures using the {DRAGONS reduction software \citep{Labrie2023}}. Flux calibration was performed with a baseline standard observation. Adopting the explosion time at $\mathrm{JD} = 2460083.25$ \citep{hosseinzadeh2023}, the phase of the new spectrum is 445 days after the explosion. This is 186 days rest-frame days after the previous spectrum of \citetalias{ferrari2024}. {Both spectra are shown in Figure~\ref{fig:nebspec}.}

\begin{figure}
    \centering
    \includegraphics[width=\linewidth]{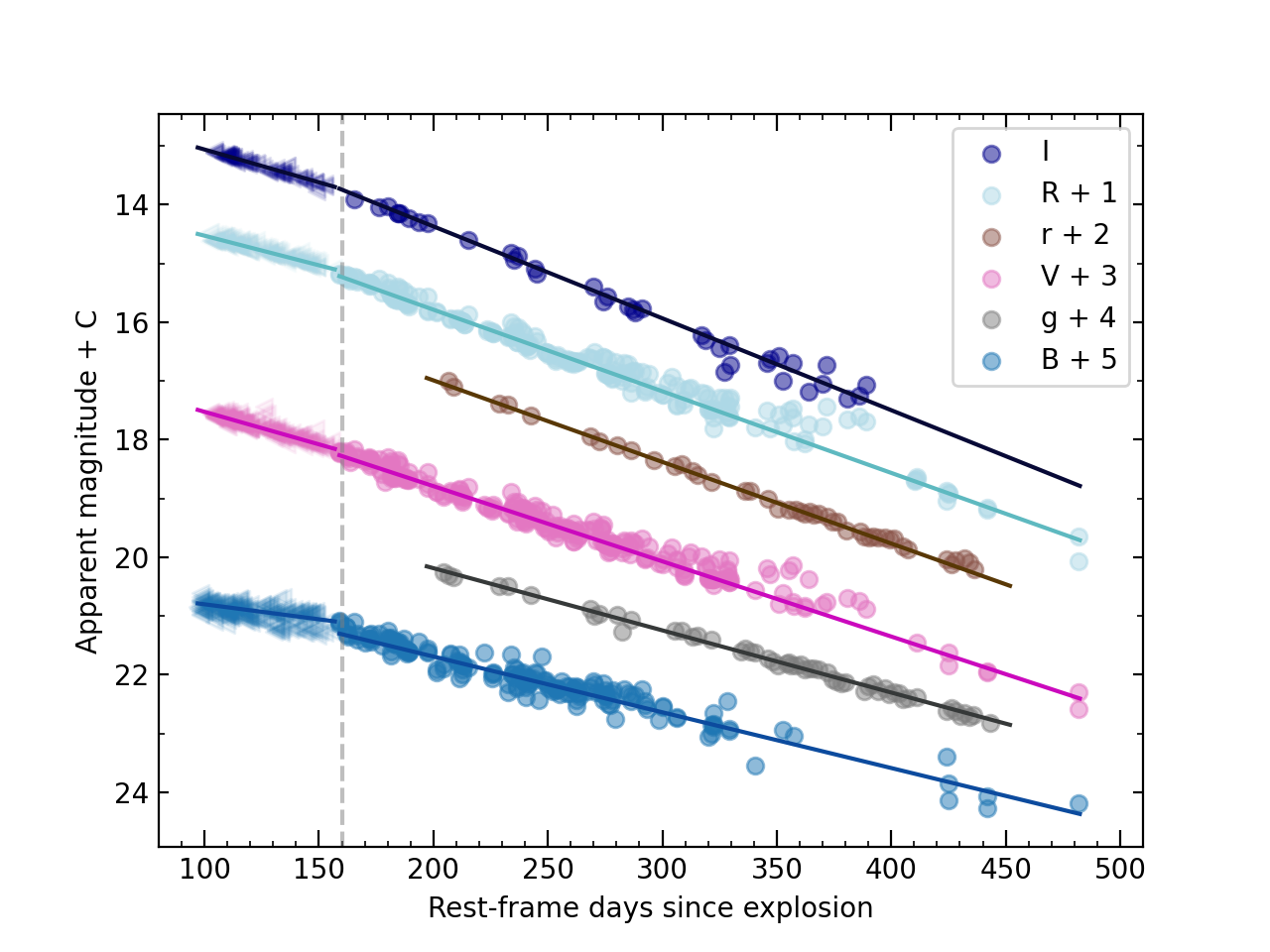}
    \caption{{Late-phase light curves of SN~2023ixf. \textit{BVRI} bands were downloaded from the AAVSO database, while \textit{g} and \textit{r} bands correspond to the ZTF program. The dashed vertical line at 160 days indicates the moment when the slopes increase in the $BVRI$ bands.}}
    \label{fig:lightcurves}
\end{figure}

Photometric $BVRI$-band data of SN~2023ixf during the nebular phase were obtained from the public database of the American Association of Variable Star Observers (AAVSO)\footnote{\url{www.aavso.org}}. The AAVSO database compiles photometry from various, mostly amateur observers around the globe. After rejecting discrepant data points, the resulting light curves contained about $500-700$ entries in the $BVR$ bands between approximately 90 and 480 days after the explosion, and 177 entries in the $I$ band between 90 and 390 days. Additional photometry in the $gr$ bands was obtained from the Zwicky Transient Facility \citep[ZTF;][]{Bellm2019,Masci2019} via the Automatic Learning for the Rapid Classification of Events broker \citep[ALeRCE;][]{Forster2021}. The ZTF photometry covers approximately the range between 200 and 480 days. {The late-time light curves described here are shown in Figure~\ref{fig:lightcurves}.}

Following \citetalias{ferrari2024}, we adopt the NASA/IPAC Extragalactic Database\footnote{\url{https://ned.ipac.caltech.edu/}} (NED) redshift of $0.0008$ for the host galaxy M101, a distance of $6.85 \pm 0.15$ Mpc \citep{Riess2022}, a Milky-Way reddening of $E(B-V)_\mathrm{MW} = 0.008$ mag \citep{Schlafly2011}, and a host-galaxy reddening of $E(B-V)_\mathrm{Host} = 0.031$ mag {\citep{Lundquist2023,Zhang2023}}. We perform extinction corrections using a total $E(B-V)_\mathrm{Total} = 0.039$ mag, and a standard extinction law by \citet{Cardelli1989} with $R_V = 3.1$.

\section{{Late-time evolution}}
\label{sec:late-ev}

\subsection{{Interaction signatures in the spectrum}}
\label{sec:inter}

Figure~\ref{fig:nebspec} shows the 445-d spectrum in comparison with that from \citetalias{ferrari2024} at 259 d. Both spectra were shifted to the SN rest frame and corrected for extinction using the redshift and extinction values provided in Section~\ref{sec:obs}. We used linear fits to the observed light curves (see Section~\ref{sec:lc_decline}) to interpolate the $VR$-band and extrapolate the $I$-band magnitudes at 445 d. By comparing these values with synthetic photometry from the spectrum, we found differences of $0.7$, $0.8$, and $0.8$ mag relative to the observed photometry in the $V$, $R$, and $I$ bands, respectively. On average, this meant a correction of $0.78$ mag or a factor of $2.05$ to be applied to the 445-d spectrum. Finally, the 259-d spectrum was scaled down to match the $R$-band magnitude interpolated at 445 days, as previously described.

We note a general agreement in the line and continuum levels between the spectra, after scaling. However, some differences are evident. First, the flux of the \ion{Ca}{ii} IR triplet feature is substantially reduced in the later spectrum (see also panel g of Figure~\ref{fig:linesvel}). This is to be expected, given that the ejecta become increasingly diluted with time and thus permitted lines lose strength. We note that a reduction of the \ion{Ca}{ii} IR triplet feature is seen in the interaction-powered model spectra in comparison with the non-interacting model, as shown by \citet[][see their Figure~4]{Dessart2023}. Therefore this decrease of the \ion{Ca}{ii} IR triplet feature may be considered as an indication of ejecta--CSM interaction. The shape of the feature also undergoes some changes, with a notable reduction of the two bluest components relative to the reddest one. 

Another distinction emerges in the region blueward of $\approx$5000\,$\AA$, where the later spectrum exhibits an augmented flux level. This is consistent with the observed slower decline of the $BV$-band light curve in comparison to the declines in the $RI$ bands, as detailed in Section~\ref{sec:lc_decline}). An excess in the blue continuum has been identified as an indicator of ejecta-CSM interaction \citep{Dessart2022}. Moreover, \citet{Dessart2023} show that the excess flux in the blue range is not attributable to continuum emission but rather to the superposition of an \ion{Fe}{i} and \ion{Fe}{ii} line forest (see their Figure~A.3). Nonetheless, a relative brightening in the blue range can also occur due to the presence of a light echo. 

Notably, as seen in Figure~\ref{fig:nebspec}, an excess flux is observed beyond 9000~\AA\ at 445~d, in comparison with the 259~d spectrum. This feature may be interpreted as a flat-topped, blueshifted, broad emission from the \ion{Mg}{ii}\,$\lambda\lambda\,9218,9244$ line that, according to \citet{Dessart2023}, may arise as a consequence of ejecta--CSM interaction. Further evidence of the interaction nature of this emission is given in Section~\ref{sec:halpha}, where we analyze the line profile.

The fourth and arguably most remarkable difference is seen in the shape of the H$\alpha$ line. The new spectrum presents a very complex profile (see Section~\ref{sec:halpha} for a detailed analysis), wherein the relatively wide profile observed on day 259 \citepalias{ferrari2024} appears to have decreased in flux and shifted further to the blue, while two additional, very wide components flank it on either side. This is possibly another indication of interaction between the SN ejecta and a dense CSM shell \citep{Dessart2023}. Unfortunately, it is not possible to test whether H$\beta$ suffers the same evolution because it is likely embedded in a superposition of mostly \ion{Fe}{i} and \ion{Fe}{ii} lines, as mentioned above.

This type of evolution of the H$\alpha$ profile has been seen in other SNe~II, as well as in some Type IIb SNe, such as SN~1993J \citep{Filippenko1994,Matheson2000}, and SN~2013df \citep{Maeda2015}. Boxy H$\alpha$ profiles have been observed in SNe~II to appear in otherwise spectroscopically and photometrically normal events, such as SNe~2004et \citep{Kotak2009}, 2007it \citep{Andrews2011}, 2008jb \citep{Prieto2012}, 2013ej \citep{Mauerhan2017}, and 2017eaw \citep{Weil2020}. However, these observations occurred at later epochs (after $\approx$700 days) than in SN~2023ixf. The lack of spectral coverage at $\sim$1000 days or after in the majority of events precludes an assessment of whether all SNe~II eventually transition to spectra dominated by interaction. Interestingly, other SNe~II that showed multiple peaks or boxy emission starting as early as $200 - 400$ days were luminous, linear or short-plateau events. Some examples are SNe~1979C \citep{Branch1981}, 2007od \citep{Inserra2011}, 2011ja \citep{Andrews2016}, 2013by \citep{Black2017}, 2014G \citep{Terreran2016}, 2017ahn \citep{Tartaglia2021}, 2017gmr \citep{Andrews2019}, 2017ivv \citep{Gutierrez2020}. With a plateau duration of $\approx$80 days \citep{bersten2024}, which is relatively short compared to the distribution of durations among SNe~II \citep{Martinez2024}, SN~2023ixf falls well in this latter group of early-interacting events.  

\begin{figure*}
    \centering
    \includegraphics[width=.9\linewidth]{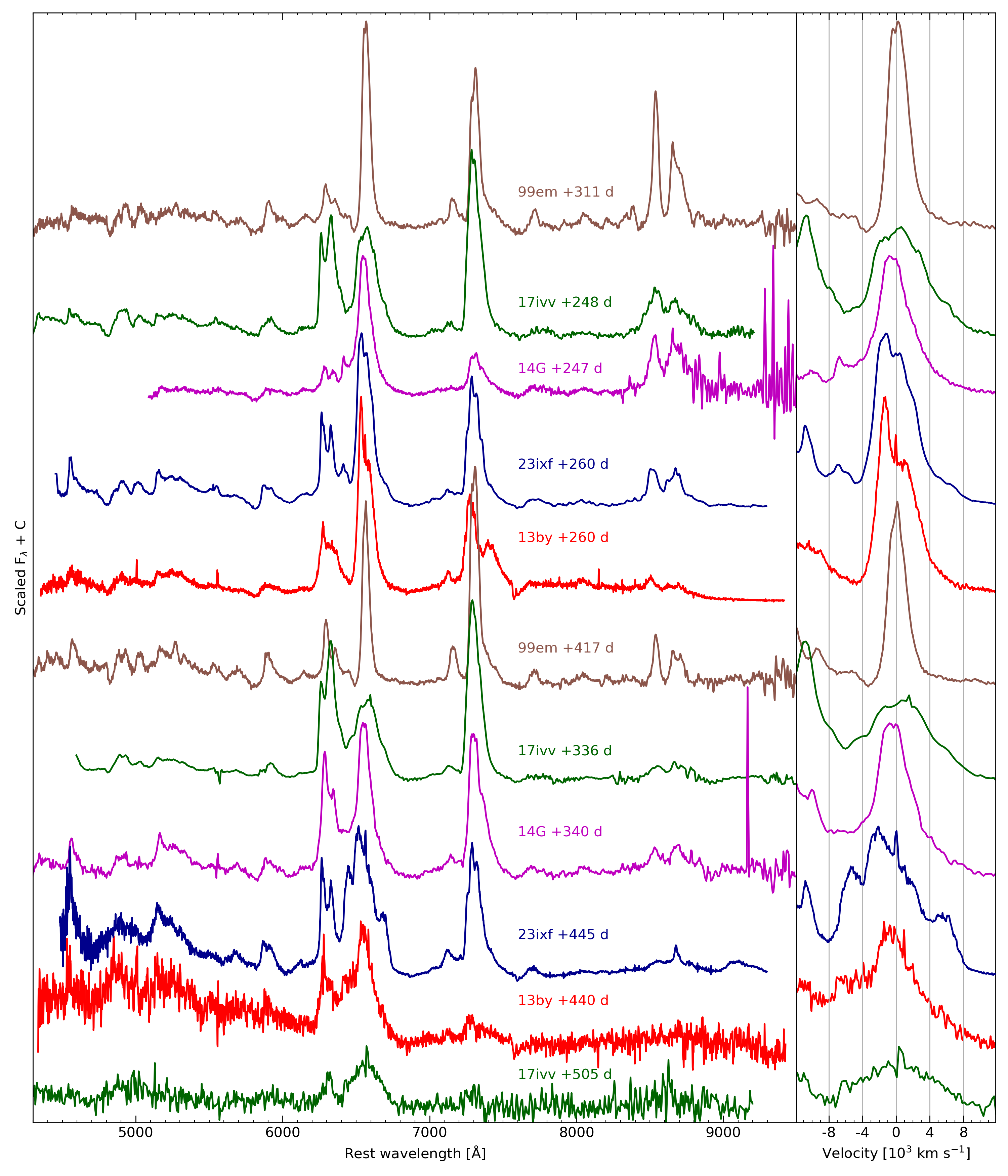}
    \caption{Nebular spectra of SN~2023ixf at 259 and 445 days (blue lines) in comparison with spectra of other SNe~II that exhibited interaction signatures at similar ages: SN~2013by \citep{Black2017}, SN~2014G \citep{Terreran2016}, and SN~2017ivv \citep{Gutierrez2020}. For reference, spectra of the prototypical SN~1999em \citep{Faran2014} are shown where no widening of the H$\alpha$ profile was observed. The right panel shows a detail of the H$\alpha$ profiles as a function of velocity relative to the rest wavelength of 6563~\AA. Comparison data were downloaded from the Wiserep \citep{yaron2012} database (\url{https://www.wiserep.org/}) and were not corrected for extinction.}
    \label{fig:neb_comp}
\end{figure*}

Figure~\ref{fig:neb_comp} shows our nebular spectra of SN~2023ixf and some similar events, along with the prototypical Type II-P SN~1999em. Possibly the closest match to the [\ion{O}{i}]\,$\lambda\lambda\,6300,6364$ and H$\alpha$ profiles at around 450 days is that of SN~2013by, although the extent of the red wing is larger in SN~2023ixf. Notably, however, the [\ion{Ca}{ii}]\,$\lambda\lambda\,7291,7324$ line is substantially stronger in SN~2023ixf. At the epoch of our previous spectrum ($\approx$260~d), the [\ion{Ca}{ii}] emission is similarly strong in both SNe, although with a complex profile in SN~2013by, possibly due to an asymmetric ejecta distribution \citep{Black2017}\footnote{Note that there can be contamination from adjacent [\ion{Fe}{ii}] and [\ion{Ni}{ii}] lines.}. Another similar event, SN~2017ivv, showed a stark reduction in the [\ion{Ca}{ii}] strength between $\approx$340~d and $\approx$500~d. 

The spectroscopic evolution of SN~2023ixf between $\approx$250~d and $\approx$450~d qualitatively follows what is shown by the radiative transfer models by \citet{Dessart2023}. The interaction signatures appear slightly earlier in SN~2023ixf than in the models. From Figure~4 (left panel) of that work it can be seen that the 445-d spectrum presented here is most similar to the model spectrum at 600~d. This phase appears to represent a midpoint in the transition between decay-power dominated and interaction-power dominated spectra. The transition is evident by the simultaneous presence of a narrower, peaked component and a broader, boxy component of H$\alpha$ (see Section~\ref{sec:halpha}).

\subsection{Line profiles}
\label{sec:line_profiles}

Figure~\ref{fig:linesvel} shows the profiles as a function of velocity for the main emission lines in the 259-d and 445-d spectra. Strikingly, the shapes of the [\ion{O}{i}]\,$\lambda\lambda\,6300,6364$ and [\ion{Ca}{ii}]\,$\lambda\lambda\,7291,7324$ features remain nearly unchanged {(see panels c) and f), respectively). The same appears to be the case of the \ion{Na}{i}\,D line (panel b) that exhibits a P-Cygni profile of similar strength at both epochs. The observed change in this feature can be attributed to the continuum level.}

{Another feature that suffers little evolution is the bump observed at around 7700~\AA (see Figure~\ref{fig:nebspec}). Its identification is unclear. It may be attributed to \ion{O}{i}\,$\lambda\,7774$, although the region may be strongly affected by \ion{Fe}{i} and \ion{Fe}{ii} emission lines \citep{Dessart2023}. If this feature were indeed due to \ion{O}{i}\,$\lambda\,7774$, its profile would differ from that of [\ion{O}{i}]\,$\lambda\lambda\,6300,6364$, displaying a single peak centered at $\approx$$-3000$~km~s$^{-1}$. A possible blueshifted absorption, such as those of P-Cygni profiles, may also be present at both epochs. However we note that this region may be affected by uncorrected telluric absorption.}

{During the same time, other lines, such as \ion{Mg}{i}]\,$\lambda\,4571$ (panel a) and [\ion{Fe}{ii}]\,$\lambda\lambda\,7155,7171$ (panel e) exhibit an increase in flux relative to the continuum. Nevertheless, the shapes of these two features do not evolve substantially and their peaks show a rather constant blueshift of $\approx$1000-1200~km~s$^{-1}$.}

{Remarkably, the most obvious changes in the shape of the line profiles are seen in H$\alpha$ (panel d), \ion{Mg}{ii}\,$\lambda\lambda\,9218,9244$ (panel h), and \ion{Ca}{ii}\,$\lambda\lambda\,8498,8542,8662$ (panel g).} 

\subsubsection{{Unchanged [\ion{O}{i}] and [\ion{Ca}{ii}] profiles}}
\label{sec:oicaii}

The [\ion{O}{i}]\,$\lambda\lambda\,6300,6364$ feature exhibits a double-peaked profile that can be well represented by two Gaussian components (see Figure~\ref{fig:OIHafit}). Both components have a width (FWHM) of $\approx$2100~km~s$^{-1}$ and are centered at $-1250$ and $+1500$~km~$^{-1}$ relative to the location of the [\ion{O}{i}]\,$\lambda\,6300$ line. The width and location of these lines are similar to those invoked by \citetalias{ferrari2024} to fit the 259-d spectrum. In that previous work the [\ion{O}{i}] fit included two extra narrow (FWHM\,$=500$~km~s$^{-1}$) components that might be present in the 445-d spectrum as well. However, the lower signal-to-noise ratio at 445 days prevents us from confirming this. We also note that no attempt is made in our fitting to reproduce the "bridge" emission seen between [\ion{O}{i}] and H$\alpha$. An interpretation of such emission will be given below in our analysis of the H$\alpha$ profile. The lack of evolution in the double-peaked [\ion{O}{i}] profile suggests the same asymmetric geometry for the oxygen-rich region of the ejecta as was interpreted by \citetalias{ferrari2024}. 

As shown in Figure~\ref{fig:linesvel}, the [\ion{Ca}{ii}]\,$\lambda\lambda\,7291,7324$ profile is remarkably similar at 445 and 259 days. Both its equivalent width and the specific structure of peaks and shoulders remain nearly unchanged. This is suggestive that we are seeing the same Ca-rich emitting region at both epochs. From a single Gaussian profile fit we derive a width of FWHM\,$=4100$~km~s$^{-1}$.

\subsubsection{{The transformation of H$\alpha$}}
\label{sec:halpha}

The shape of the H$\alpha$ line is where most of the evolution is seen between 259 and 445 days {(see Figure~\ref{fig:nebspec})}. At the earlier epoch, the profile could be interpreted as composed by a broad ($\mathrm{FWHM}=5300$~km~s$^{-1}$) emission centered at rest plus a narrower ($\mathrm{FWHM}=1600$~km~s$^{-1}$) component blue-shifted by 1800~km~s$^{-1}$ \citepalias{ferrari2024}. At 445 days, there appears to be a central broad emission with a slanted top profile superimposed on an even broader structure {with sharp edges}. This complex profile can be fit with multiple Gaussian components, as shown in Figure~\ref{fig:OIHafit}. The fit was done simultaneously for H$\alpha$ and [\ion{O}{i}]\,$\lambda\lambda\,6300,6364$ due to the blending between both features. 

\begin{figure*}
    \centering
    \includegraphics[width=\linewidth]{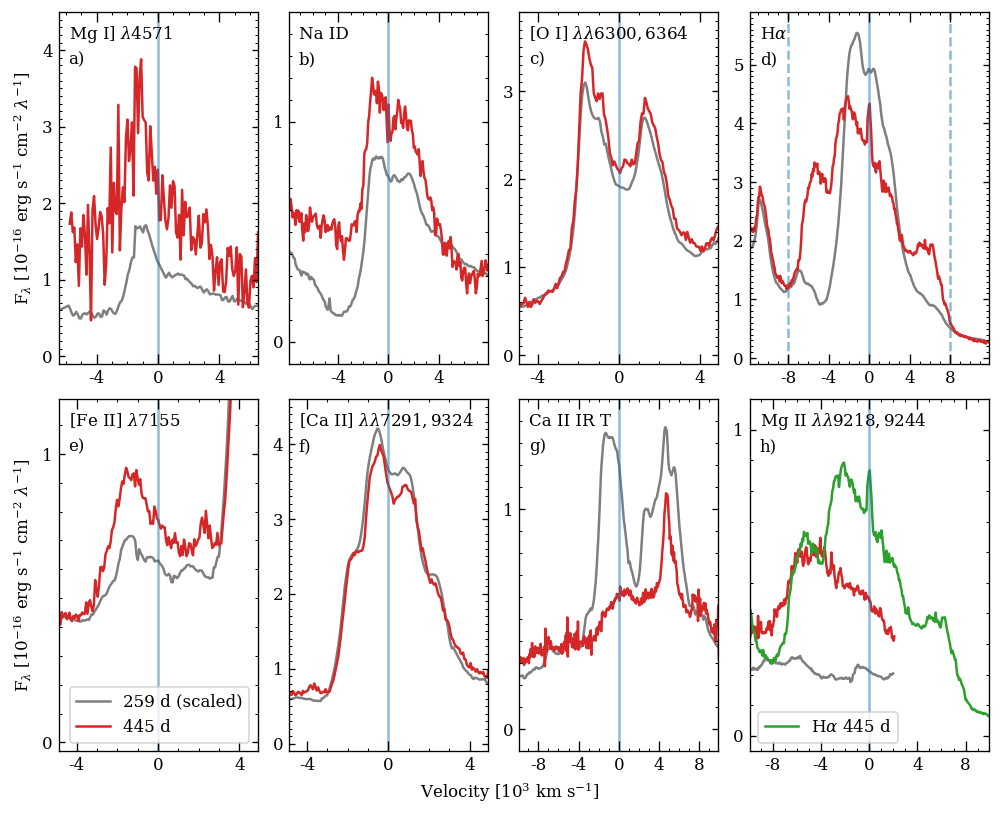}    
    \caption{Line profiles as a function of velocity for the main emission features in the spectra of SN~2023ixf at 259 days (gray lines) and 445 days (red lines). Similarly to Figure~\ref{fig:nebspec}, both spectra were scaled to match the $R$-band photometry at 445 days. Vertical lines indicate the location of rest (solid) for the reference wavelength of each line, as given below, and in panel d), additionally at velocities of $\pm$8000~km~s$^{-1}$ (dashed). Panel h) shows the region of the \ion{Mg}{ii}\,$\lambda\lambda\,9218,9244$ feature, which is only partly covered, and it includes for reference a reproduction of the H$\alpha$ profile at 445 days (green line). Reference wavelengths in each panel are: a) 4571~\AA, b) 5890~\AA, c) 6300~\AA, d) 6563~\AA, e) 7155~\AA, f) 7300~\AA, g) 8542~\AA, and h) 9231~\AA.}
    \label{fig:linesvel}
\end{figure*}

\begin{figure}
    \centering
    \includegraphics[width=\linewidth]{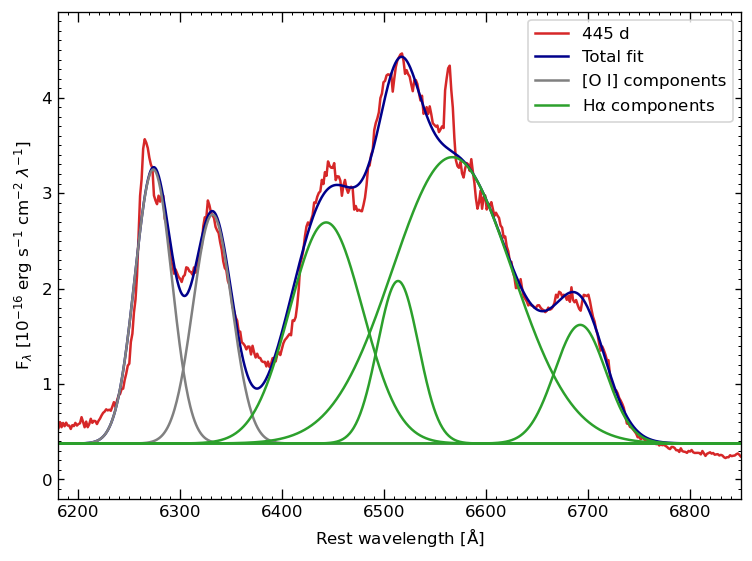}
    \caption{Multi-component Gaussian fit to the [\ion{O}{i}]\,$\lambda\lambda\,6300,6364$ and H$\alpha$ emission complex in the 445-d spectrum of SN~2023ixf (red line). The sum of all six Gaussian components that fits the observations is shown with a blue line. Two components are associated with the [\ion{O}{i}]\,$\lambda\lambda\,6300,6364$ feature (gray lines), while other four components reproduce the H$\alpha$ profile (green lines). See the text for further explanations.}
    \label{fig:OIHafit}
\end{figure}

We achieve a reasonable match to the observed H$\alpha$ profile by invoking four Gaussian components. {The number of Gaussians was chosen in order to reproduce the two-Gaussian fit performed by \citetalias{ferrari2024} on the 259-d spectrum plus two additional components at larger displacements on each side that reproduce the broader extent of the line.} {The widest of these components, with $\mathrm{FWHM}=6300$~km~s$^{-1}$, is centered near the rest wavelength at $\approx$$+150$~km~s$^{-1}$. An additional narrower component, with $\mathrm{FWHM}=2200$~km~s$^{-1}$ and blueshifted by $\approx$$-2200$~km~s$^{-1}$, accounts for the peak of the central emission. Both of these central components are slightly wider than those fit to the 259-d spectrum. Finally, two roughly symmetrical Gaussians centered at $-5500$ and $+5900$~km~s$^{-1}$ and with widths of FWHM\,$\approx$\,3900 and 2700~km~s$^{-1}$, respectively, reproduce the extreme of the wider boxy emission.} The Gaussian on the blue extreme has a larger amplitude than its counterpart on the red end. We note that this multi-component fit is slightly wider than the observations on the blue edge of the H$\alpha$ profile. This may be due the presence of an extra emission component in between the [\ion{O}{i}]\,$\lambda\lambda\,6300,6364$ line and H$\alpha$ that is not accounted for in the fit. {Blending with the [\ion{O}{i}] feature may also explain the larger FWHM value found for the bluest Gaussian as compared with that of the reddest component in the H$\alpha$ fit.} In addition, the fit does not reproduce an apparent depression in the spectrum at about 6470~\AA. This depression may be due to an absorption component of H$\alpha$ centered at about $-4250$~km~s$^{-1}$. A similar feature is noticeable in the 259-d spectrum although at a higher velocity (see panel d of Figure~\ref{fig:linesvel}).

{The two extreme Gaussian components may in principle arise from a bipolar, high-velocity ejecta distribution. However, if we assume them to be powered by $^{56}$Co decay, these high-velocity emission components should have been present at phases earlier than 259 days, which is not the case \citep[see also][]{Kumar2025,Michel2025}. Their late-phase appearance is instead likely associated with the fast SN ejecta encountering pre-existing, slowly-moving CSM. Under this interpretation, the central components may be dominated by emission from the SN ejecta\footnote{{Note, however, that this central emission may also be affected by interaction with an asymmetric CSM \citep[see][]{Kumar2025}}.},} as its width is comparable to that seen at 259 days, whereas the broader boxy component may be the result of interaction between the SN ejecta and a pre-existing CSM that started some time between 259 and 445 days. The slanted top of the central component may be due in this scenario to obscuration of the far side of the ejecta, which would explain the reduced flux on the red side of the line and the blueshift of its peak. A similar type of obscuration may affect the red side of the broader component, which is weaker than the blue side. The sharp edges seen on each side at $\approx$$\pm$8000~km~s$^{-1}$ may be produced by the emission from a roughly {spherical shell or axisymmetrical structure} where the shock front is located. 

We note that while H$\alpha$ exhibits this transformation and its central peak shifts to the blue, the peaks of the other spectral features remain nearly fixed in place (see Figure~\ref{fig:linesvel}). This may be explained if the obscuration region partially covers the outer H-emission layers thus preferentially allowing (blueshited) photons from the near side of the ejecta to escape \citep{Chugai1992}. Such a phenomenon may occur if the obscuration region resides in a complex, asymmetric or clumpy structure that might be associated with the ejecta--CSM interaction front. In turn, lines from heavier elements would form deeper in the ejecta where the obscuration zone would equally absorb photons from the near (blueshifted) and far (redshifted) emitting regions thereby keeping their central wavelengths unchanged.

{Following the above scenario, we present an alternative representation of the H$\alpha$ profile as the sum of an underlying boxy component, in the form of a trapezoid, and a blueshifted, more centrally peaked component}. Assuming that the red side of the line is affected by absorption, we use only the data blueward of the rest wavelength to produce a fit. The trapezoid is fit to the blue extreme of the profile, i.e. the regions between $\approx$$-4000$ and $-8000$~km~s$^{-1}$. For the peaked component we fit the data from 0 to $\approx$$-4000$~km~s$^{-1}$ using two half-Gaussians, one in emission and one in absorption. The choice of a Gaussian shape for the absorption component is arbitrary and it was made for simplicity. We noted that the central Gaussian was poorly constrained given the relatively narrow and steep profile. Therefore, we fixed the amplitude of the positive Gaussian to match that of the 
H$\alpha$ emission in the scaled-down 259-d spectrum, assuming a similar origin for the feature.

\begin{figure}
    \centering
    \includegraphics[width=\linewidth]{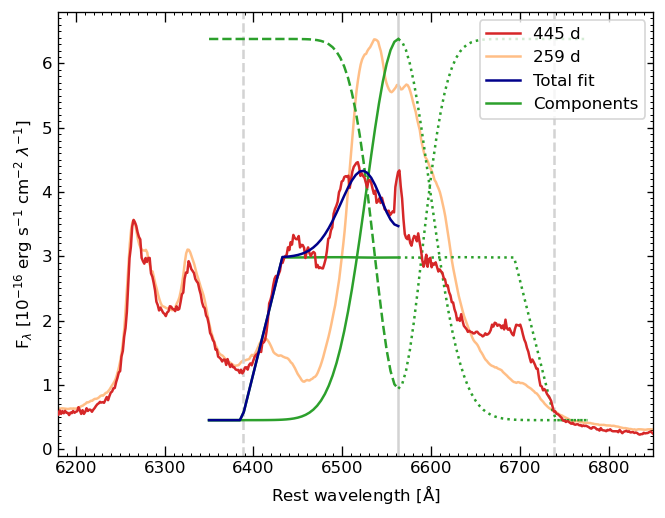}
    \caption{Simulated interaction-powered (trapezoid) plus radoiactive-powered (Gaussian) fit to the blue side of the H$\alpha$ profile in the 445-d spectrum of SN~2023ixf (red line). The total fitted profile is shown with a blue line. Green lines show the individual fit components: trapezoid and emission Gaussian in solid line, and absorption Gaussian in dashed line. Green dotted lines indicate the symmetrical extension of each component toward the red half of the profile. Note that the trapezoid matches well the observed red edge of the line. Gray vertical lines indicate the location of rest (solid) and $\pm8000$~km~s$^{-1}$ (dashed). 
    The scaled-down 259-d spectrum of SN~2023ixf \citepalias[orange line,][]{ferrari2024} was used to fix the amplitude of the emission Gaussian. See the text for further details.}
    \label{fig:Haespejo}
\end{figure}

Figure~\ref{fig:Haespejo} shows the result of simultaneously fitting the trapezoid and the positive and negative Gaussians to the blue side of the H$\alpha$ line.
No attempt was made to reproduce the dip at $\approx$6470~\AA\ mentioned above. {The dotted lines in the Figure show the three individual fit components mirrored toward the red side of the line.} It is worth noting that the edge of the trapezoid that was fit on the blue extreme of the line matches very well the red edge, which provides support to its interpretation as the emission from a symmetrical interaction {region.} 

Within this {representation} we can derive a total unattenuated flux of the interaction-powered H$\alpha$ emission as the integrated flux of the trapezoid mirrored around the rest wavelength. Adopting the distance to M101 given in Section~\ref{sec:obs} this gives a luminosity of $4.2\times10^{38}$~erg~s$^{-1}$. This value is slightly higher than that of SN~2017eaw at 900 days, and about one order of magnitude larger than what was measured on SN~2004et and SN~2013ej at similar ages when they were undergoing interaction \citep[see][]{Weil2020}. Similar H$\alpha$ interaction-driven luminosities of $\sim$$10^{38}$~erg~s$^{-1}$ were found for the Type IIb SN~1993J at 367 d \citep{Patat1995}, and SN~2013df at 626 d \citep{Maeda2015}. 

The unattenuated H$\alpha$ luminosity due to the ejecta in turn can be considered as that of the emission Gaussian, which is $3.0\times10^{38}$~erg~s$^{-1}$. This value is also comparable with those of non-interacting SNe~II at a similar age \citep{Weil2020}. However, this assumed radioactive-powered H$\alpha$ luminosity is not well-constrained for SN~2023ixf because we fixed the line strength based on the scaled-down 259-d spectrum.

Panel h) of Figure~\ref{fig:linesvel} depicts the profile evolution of the feature we identified as \ion{Mg}{ii}\,$\lambda\lambda\,9218,9244$. Unfortunately, the profile is not completely covered. However, the blue side exhibits a "boxy" profile, similar to that of H$\alpha$, extending to $\approx$$-8000$~\AA. The top of the emission appears slanted toward the blue side, also in agreement with the H$\alpha$ profile. If this identification is correct, it suggests that \ion{Mg}{ii} is the species that most closely follows the behavior of the interaction-driven H$\alpha$ emission. 

It is interesting to examine the evolution of the region in between H$\alpha$ and the [\ion{O}{i}] feature at around 6400~\AA. At 259 days the spectrum exhibits a peak centered at 6412~\AA\ that is not usually seen in SN-II spectra \citep{Silverman2017}. \citetalias{ferrari2024} suggested this emission could be due to \ion{Fe}{ii}\,$\lambda\,6456$ or \ion{O}{i}\,$\lambda\,6456$ blueshifted by $\approx$$-2000$~km~s$^{-1}$. A similar feature appeared in the spectra of SN~2014G between 100 and 190 days that was possibly identified as high-velocity H$\alpha$ by \citet{Terreran2016}. If the identification is correct, the velocity of the emission peak evolved in that SN from $\approx$$-7600$ to $\approx$$-6800$~km~s$^{-1}$ during the time that it was detected. We further notice a similar emission feature centered at $\approx$$-7000$~km~s$^{-1}$ in the spectra of SN~2017ivv between 242 and 285 days \citep{Gutierrez2020}, much alike the case of SN~2023ixf. Interestingly, as can be appreciated in Figure~\ref{fig:linesvel}, the location of the peak at 259 days roughly matches the blue edge of the boxy H$\alpha$ line at 445 days. Blueward of the emission peak, the spectrum shape remains unchanged between the two epochs. This may indicate a common origin for both features, which would support the H identification of the peak seen at 259 days. 

It is unclear whether this possible H feature is related with high-velocity "cachito" absorption features commonly observed during the plateau phase \citep{Gutierrez2017}, which have been attributed to interaction between the SN ejecta and the RSG progenitor's wind \citep{Chugai2007}. High-velocity components of H$\alpha$ and H$\beta$ were seen in SN~2023ixf during the plateau phase \citep{Singh2024}, although the authors discuss that they are more likely due to an asymmetric ejecta rather than to CSM interaction. According to the radiation transfer calculations of SNe~II with CSM interaction by \citet{Dessart2022} and \citet{Dessart2023}, this type of absorption is persistent well into the nebular phase. However, we cannot identify it in the late-time spectra of SN~2023ixf.

We can compute the internal radius of a putative CSM shell {(or distance to the CSM structure)} by assuming the SN ejecta traveling at the maximum speed of 8000~km~s$^{-1}$ given by the H$\alpha$ profile, and the interaction starting at some time between 259 and 445 days since the explosion. This gives a inner radius in the range of $1.8$ to $3.0\times10^{16}$~cm. Assuming that the CSM shell was produced by a wind expelled from the RSG progenitor at a velocity of $\sim$10~km~s$^{-1}$, then this material must have been lost from the star up until $\sim$$500 - 1000$ years prior to the SN event. If we consider the emission peak seen at $\approx$6400~\AA\ in the 259-d spectrum as due to CSM interaction, then the lowest radius and age values above are to be taken as upper limits. 

\subsection{Light-curve decline}
\label{sec:lc_decline}

The extra power released by the ejecta--CSM interaction can produce a flattening of the light curves. This effect has been observed as a break in the decline slopes for a number of SNe~II following the appearance of the boxy H$\alpha$ profiles. Notably, the break occurred earlier (before 500~d) in SN~2007od \citep{Andrews2010,Inserra2011}, SN~2011ja \citep{Andrews2016}, and SN~2017ivv \citep{Gutierrez2020} than in SN~2004et \citep[after 600~d;][]{Maguire2010}, SN~2013ej \citep[possibly after 750~d;][]{Mauerhan2017}\footnote{Note, however, that \citet{Dhungana2016} found a break in the light curves of SN~2013ej at $\approx$180 days.}, and SN~2017eaw \citep[after 700~d;][]{Weil2020}, roughly matching the timing of the spectral interaction signatures (see Section~\ref{sec:inter}). This supports the association of both spectroscopic and photometric effects to the same physical cause, most probably CSM interaction. Lack of sufficient coverage in the light curves of other SNe~II with interaction signatures in the spectra prevents us from driving a strong general conclusion about this association. In addition, we note that \citet{Andrews2010} suggested the flattening of the light curves in SN~2007od was due to a light echo from dust that in turn obscured the far side of the ejecta causing asymmetric line profiles in the late spectra. 

{Figure~\ref{fig:lightcurves} shows the late-time optical light curves of SN~2023ixf (see Section~\ref{sec:obs}) between $\approx$100 and $\approx$480 days. During this time there is no evidence of a decrease in the decline rates. Moreover, \citet{Michel2025} found no apparent break in the late-time optical light curves until $\approx$520 d, with the possible exception of the $I$ and $z$ bands, which the authors suggest to be associated with emission from pre-existing or newly-formed dust.}

Since the onset of the radioactive tail at $\approx$90 days, the $BVRI$ light curves show a relatively slow decline until $\approx$160 days {(dashed line in Figure \ref{fig:lightcurves})}. After that, the slopes increase and remain roughly constant in all bands until the end of the coverage ($\approx$480 days in $BVR$, and $\approx$390 days in $I$ band). We perform linear fits to the photometric data during both epoch ranges. In the range of 90 -- 160~d, the slopes are $0.51 \pm 0.06$, $1.11 \pm 0.04$, $1.01 \pm 0.13$, and $1.12 \pm 0.09$~mag (100~d)$^{-1}$ in $BVRI$, respectively. At times later than 160~d, the slopes in the same bands are $0.95 \pm 0.06$, $1.28 \pm 0.04$, $1.39\pm 0.03$, $1.56 \pm 0.01$~mag (100~d)$^{-1}$. Note that the increase in decline slope after 160~d is substantially larger in $B$ band than in $VRI$. In addition, during both time regimes, the decline slopes are the larger the longer the filter's wavelength. The same is seen in the $gr$-band light curves from ZTF that cover the range of 200 to 444 days. The slopes are $1.06 \pm 0.01$ and $1.39 \pm 0.03$ mag (100~d)$^{-1}$ in $g$ and $r$, respectively. {The lack of change in slopes between $\approx$200 and $\approx$500 days indicates that if CSM interaction were responsible for the spectral evolution described in Sections~\ref{sec:inter} and \ref{sec:halpha}, it would not be strong enough to produce a notable change in the light curves, at least during the time covered by the data.} 

Except in the $B$ band, the decline rates calculated above are larger than the rate of $0.98$~mag (100~d)$^{-1}$ that is expected from the radioactive decay of $^{56}$Co. This has been commonly seen in SNe~II by \citet{Anderson2014}, who derived an average $V$-band decline rate during the radioactive tail of $1.47 \pm 0.82$ mag (100~d)$^{-1}$. Furthermore, the authors note that SNe~II with steeper $V$-band declines during the plateau phase tend to show larger decline rates during the radioactive tail. Note also that \citet{bersten2024} measured larger than average decline rates in the bolometric light curve of SN~2023ixf both during the plateau and the radioactive tail. Similarly steep declines have been measured at $200 - 400$ days in other SNe~II with signs of interaction in the spectra at relatively early times during the nebular phase, namely SN~2013by \citep{Black2017}, SN~2014G \citep{Terreran2016}, and SN~2017ivv \citep{Gutierrez2020}, as well as in other objects with later signs of interaction in the spectra, such as SN~2004et \citep{Maguire2010}, SN~2008jb \citep{Prieto2012}, SN~2013ej \citep{Dhungana2016}, and SN~2017eaw \citep{Buta2019}.

Assuming the late-time light curve is powered solely by radioactive decay, the deviation from the expected decline rate can be due to incomplete gamma-ray trapping in the ejecta or to increased extinction by newly-formed dust, or a combination of both. Dust extinction should, however, produce systematically steeper light curves in the bluer bands, which is contrary to what was observed.

\section{Progenitor mass revisited}

\label{sec:prog_mass}

Nebular spectroscopy can be used to estimate the initial mass of the progenitor. One way to do this is by deriving the mass of oxygen in the ejecta and linking that mass with oxygen yields from explosion models of stars with different initial masses. The mass of neutral oxygen can be determined from the flux of the [\ion{O}{i}]\,$\lambda\lambda\,6300,6364$ line, provided that the temperature of the emitting gas is known \citep{Uomoto1986,Jerkstrand2014}. This \ion{O}{i} mass is in principle a lower limit to the total oxygen mass in the ejecta, bearing the fraction of ionized oxygen \citep{Kuncarayakti2015}. The temperature can be estimated by measuring the flux ratio of [\ion{O}{i}]\,$\lambda\,5577$ to [\ion{O}{i}]\,$\lambda\lambda\,6300,6364$ \citep[see equation 2 in][]{Jerkstrand2014}. However, [\ion{O}{i}]\,$\lambda\,5577$ is a weak line and in many cases it is lost in the noise. Moreover, the dependence of the \ion{O}{i} mass on temperature is through an exponential, which makes it very sensitive to uncertainties in the temperature \citep[see][]{Elmhamdi2004}. 

\citetalias{ferrari2024} placed a lower limit on the [\ion{O}{i}]\,$\lambda\,5577$ flux in the 259-d spectrum of SN~2023ixf, and thereby derived a lower limit for the temperature of 3700~K. Based on the [\ion{O}{i}]\,$\lambda\lambda\,6300,6364$ flux of $2.38\times\,10^{-13}$~erg~s$^{-1}$~cm$^{-2}$ measured on the 259-d spectrum, the authors found an \ion{O}{i} mass of $\approx\,0.5$~M$\odot$. The [\ion{O}{i}]\,$\lambda\lambda\,6300,6364$ flux at 445 days can be obtained from the Gaussian fits described in Section~\ref{sec:oicaii}. The result is $1.2\times\,10^{-14}$~erg~s$^{-1}$~cm$^{-2}$, i.e., 20 times smaller than that at 259 days. The [\ion{O}{i}]\,$\lambda\,5577$ line is undetectable at 445 days, which prevents a temperature estimate to be obtained. Assuming a temperature of 3330~K measured from the 451-d spectrum of SN~2012aw by \citet{Jerkstrand2014}, the resulting \ion{O}{i} mass from the 445-d spectrum of SN~2023ixf would be $0.05$~M$\odot$. This is a factor of ten lower than the mass derived from the 259-d spectrum. In order to recover the same \ion{O}{i} mass at 445 days as at 259 days, and assuming all other physical conditions remain constant, the required temperature at 445 days would be $\approx$2500~K. Given the amount of assumptions and the critical dependence on temperature involved in this method, we move on to directly compare the observed spectrum with synthetic models. 

\begin{figure}
    \centering
    \includegraphics[width=\linewidth]{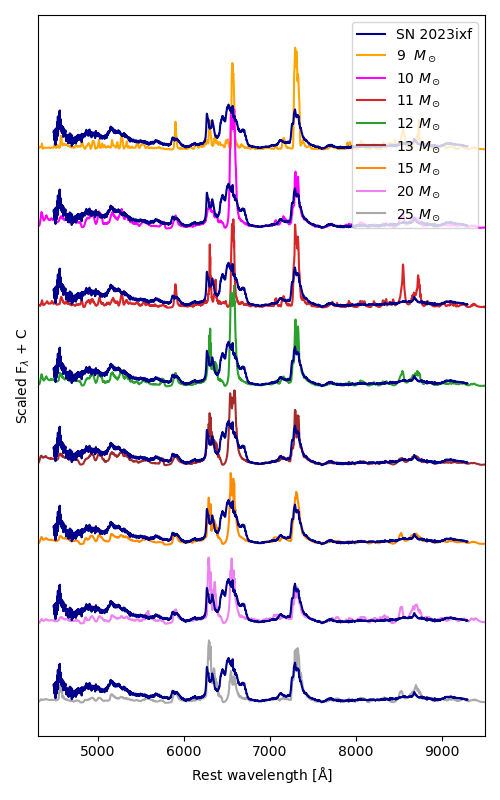}
    \caption{The spectrum of SN~2023ixf at 445 days (blue lines) in comparison with radiative transfer calculations at an age of 360 days performed by \citet{Dessart2021} for various SN~II progenitors of different masses (colored lines as labeled). The model spectra were normalized to match the observed flux integrated between 6900 and 8200~\AA.}
    \label{fig:compdessart21}
\end{figure}

\begin{figure}
    \centering
    \includegraphics[width=\linewidth]{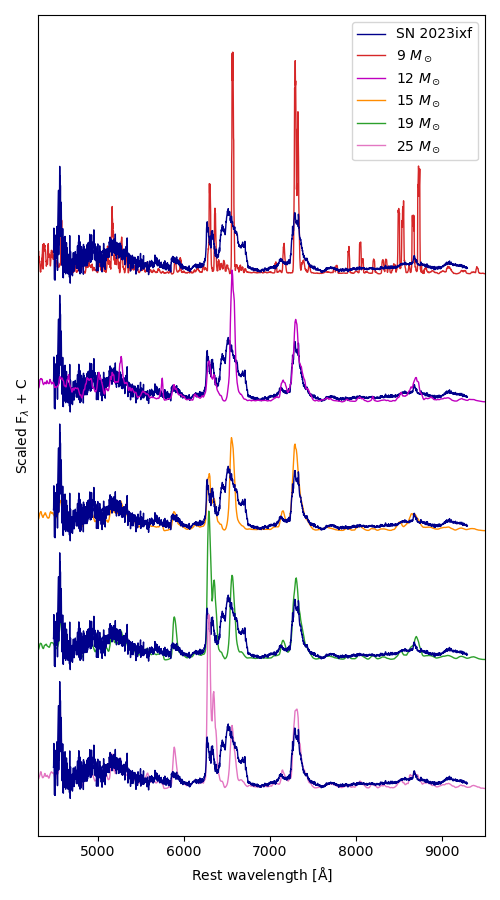}
    \caption{Similar to Figure~\ref{fig:compdessart21} but the 445-d spectrum (blue lines) is compared with radiative transfer calculations of SN~II spectra at 450 days for progenitors of various initial masses (colored lines as labeled) computed by \citet{Jerkstrand2018} for 9~M$_\odot$, and \citet{Jerkstrand2014} for the rest of the masses. The model spectra were normalized to match the observed flux integrated between 6900 and 8200~\AA.}
    \label{fig:compjerkstrand}
\end{figure}

Figure~\ref{fig:compdessart21} shows the 445-d spectrum of SN~2023ixf in comparison with a series of SN~II synthetic spectra of varying progenitor masses calculated by \citet{Dessart2021} for an age of 350 days. Similarly, Figure~\ref{fig:compjerkstrand} presents a comparison with the models at 450 days computed by \citet{Jerkstrand2014} and \citet{Jerkstrand2018}. In both comparisons, the model spectra do not include any effect of CSM interaction, therefore the H$\alpha$ line and the region below $\approx$\,5500~\AA\ are not expected to match the observations. Instead, we focus the comparison on the [\ion{O}{i}]\,$\lambda\lambda\,6300,6364$ and [\ion{Ca}{ii}]\,$\lambda\lambda\,7291,7324$ emissions, which have been used as indicators of the initial mass of the progenitor \citep[e.g., see][]{Fransson1989,Dessart2021}, and are relatively free from interaction effects, as shown by \citet{Dessart2023}. The synthetic spectra are scaled to match the average flux of the photometry-corrected spectrum (see Section~\ref{sec:inter}) in the range of $6800 - 8200$~\AA, i.e., within most of the $i'$ and $I$ bands. Therefore, the scaling factor is dominated by the [\ion{Ca}{ii}]\,$\lambda\lambda\,7291,7324$ line flux. This choice of scaling factor versus using the distance, age and $^{56}$Ni mass is made to account for differences in the fraction of $\gamma$-ray escape from the ejecta between models and observation. The comparison indicates that the observed spectrum is best matched by models with initial masses between 10 and 15 M$_\odot$ from \citet{Dessart2021}, and between 12 and 15 M$_\odot$ from \citet{Jerkstrand2014}. Models with M$_\mathrm{ZAMS}=9$~M$_\odot$ exhibit comparatively narrow emission lines that do not match the observations. On the other hand, for M$_\mathrm{ZAMS} \geq 19$~M$_\odot$ the [\ion{O}{i}]\,$\lambda\lambda\,6300,6364$ emission becomes too strong in both sets of model spectra.

\begin{figure}
    \centering
    \includegraphics[width=\linewidth]{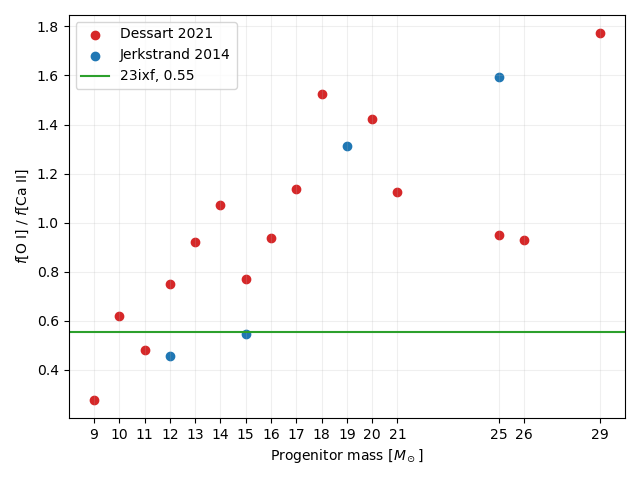}
    \caption{Flux ratio of the [\ion{O}{i}]\,$\lambda\lambda\,6300,6364$ line to the [\ion{Ca}{ii}]\,$\lambda\lambda\,7294,7321$ line as measured in the 445-d spectrum of SN~2023ixf (green horizontal line) in comparison with the ratios measured on the model spectra of various progenitor masses from \citet{Dessart2021} (red circles) and \citet{Jerkstrand2014} (blue circles). This comparison favors a progenitor mass in the range of about $10-15$~M$_\odot$.}
    \label{fig:O-Ca_ratio}
\end{figure}

A more quantitative comparison can be done using the flux ratios of [\ion{O}{i}]\,$\lambda\lambda\,6300,6364$ over [\ion{Ca}{ii}]\,$\lambda\lambda\,7291,7324$. Figure~\ref{fig:O-Ca_ratio} shows the ratios computed from both series of synthetic spectra mentioned above. {From the observed spectrum we obtain a ratio of $0.55$,} which is in agreement with that of $0.51$ obtained from the 259-d spectrum by \citetalias{ferrari2024}. This stability of the oxygen to calcium ratio has been observed in other SNe~II and in core-collapse SNe in general \citep[e.g., see][]{Elmhamdi2011,Maguire2012}, which provides support to the method. From Figure~\ref{fig:O-Ca_ratio} we can conclude that SN~2023ixf is compatible with models with initial masses between 10 and 15 M$_\odot$. 

The present analysis favors a mass of the progenitor for SN~2023ixf on the low range among those derived from the analysis of the pre-explosion imaging (see Section~\ref{sec:intro}). This is in agreement with the study of the 259-d spectrum by \citetalias{ferrari2024}, and with the hydrodynamical modelling of the light curves and expansion velocities during the plateau phase by \citet{bersten2024} and \citet{Moriya2024}.

\section{Conclusions}\label{sec:conclusion}

We have presented a second nebular spectrum of SN~2023ixf obtained with GMOS-N at 445 days after the explosion. In the approximately 180 days since our previous spectrum \citepalias[presented in][]{ferrari2024} the SN underwent a dramatic transformation. The new spectrum shows clear indications of ejecta--CSM interaction. Most notably, in the appearance of a boxy emission component of the H$\alpha$ line that may also be present in the \ion{Mg}{ii}\,$\lambda\lambda\,9218,9244$ line (see Figure~\ref{fig:linesvel}). This suggests that the ejecta encountered a CSM shell. The flux in the region blueward of $\approx$5000~\AA\ shows a relative enhancement, which has also been indicated as a sign of CSM interaction \citep{Dessart2022}. This transformation seen at 445 days\footnote{Note that \citet{Kumar2025} noticed a similar boxy H$\alpha$ profile at 363 days.} places SN~2023ixf within the group of SNe~II that have shown interaction signatures in their nebular spectra before $\approx$500 days. This group appears to be composed of SNe with either short plateaus or linear light curves. 

The boxy H$\alpha$ component extends to $\pm$8000~km~s$^{-1}$. Assuming this expansion velocity for the SN ejecta, the age of 445 days provides a distance of the putative CSM shell {(or axisymmetric structure)} from the progenitor star of $\approx$$3\times10^{16}$~cm. This is an upper limit since the interaction may have started as early as $\approx$260 days (see Section~\ref{sec:line_profiles}). In conclusion the {CSM must have been located at a distance of} $\sim$$10^{16}$~cm. Assuming a standard wind velocity for the RSG progenitor of $\sim$10~km~s$^{-1}$, the material should have been expelled by the star up until $\approx$$500-1000$ years before the explosion. We note that, contrary to what was observed in other interacting SNe~II, in SN~2023ixf the appearance of the boxy H$\alpha$ profile was not accompanied by a flattening of the light curves {at least until $\approx$500 days,} as it is expected due to the release of extra interaction power \citep{Dessart2023}. However, {it is still possible that this phenomenon may be detected at even later times.}

Some of the main emission lines in the 445-d spectrum, such as [\ion{O}{i}]\,$\lambda\lambda\,6300,6364$, [\ion{Ca}{ii}]\,$\lambda\lambda\,7291,7324$, \ion{Na}{i}~D and [\ion{Fe}{ii}]\,$\lambda\lambda\,7155,7171$, exhibited remarkably similar profiles  compared with their counterparts at 259 days (see Figure~\ref{fig:linesvel}). Assuming these features remained to be dominated by emission from the SN ejecta and not from the interaction region, this indicates that the geometry of the emitting regions of the ejecta was not substantially modified during the time elapsed between the observations.

Moreover, the fact that the [\ion{O}{i}]\,$\lambda\lambda\,6300,6364$ and [\ion{Ca}{ii}]\,$\lambda\lambda\,7291,7324$ lines remain apparently unaffected by the CSM interaction, allowed us to derive an updated estimate of the progenitor mass. This was done by comparing with synthetic nebular spectra calculated by \citet{Jerkstrand2014}, \citet{Jerkstrand2018}, and \citet{Dessart2021} for non-interacting SNe~II of various progenitor masses. The results (see Figures~\ref{fig:compdessart21}, \ref{fig:compjerkstrand}, and \ref{fig:O-Ca_ratio}) favor a progenitor mass of M$_\mathrm{ZAMS}\approx10-15$~M$_\odot$, similar to what was concluded by \citetalias{ferrari2024} using the 259-d spectrum. We note, however, that the synthetic spectra were computed assuming progenitors with standard mass-loss prescriptions. Alternatively, the progenitor may have been a star with a slightly higher initial mass but that underwent enhanced mass loss, as proposed by \citet{Fang2025}.

The proximity of SN~2023ixf allows a rare opportunity of monitoring the evolution of this event for the coming years or even decades. Future observations may reveal further details of the CSM properties and the possible formation of dust in the ejecta--CSM interface. This will provide important clues about the evolution of massive stars and specifically their mass-loss processes.

\begin{acknowledgements}
    We are thankful to Jennifer Andrews for her willingness to share data and information, and to the Gemini Observatory for encouraging fluid group interaction between the Program GN-2024A-Q-139 and Program GN-2024A-Q-309 teams. H.K. was funded by the Research Council of Finland projects 324504, 328898, and 353019. K.M. acknowledges support from JSPS KAKENHI grants JP24H01810 and JP24KK0070. K.M. and H.K. are partly supported by the JSPS bilateral program between Japan and Finland (JPJSBP120229923).
    Based on observations obtained at the international Gemini Observatory (GN-2024A-Q-309, PI: Ferrari), a program of NSF's NOIRLab, which is managed by the Association of Universities for Research in Astronomy (AURA) under a cooperative agreement with the National Science Foundation. On behalf of the Gemini Observatory partnership: the National Science Foundation (United States), National Research Council (Canada), Agencia Nacional de Investigaci\'{o}n y Desarrollo (Chile), Ministerio de Ciencia, Tecnolog\'{i}a e Innovaci\'{o}n (Argentina), Minist\'{e}rio da Ci\^{e}ncia, Tecnologia, Inova\c{c}\~{o}es e Comunica\c{c}\~{o}es (Brazil), and Korea Astronomy and Space Science Institute (Republic of Korea). This work was enabled by observations made from the Gemini North telescope, located within the Maunakea Science Reserve and adjacent to the summit of Maunakea. We are grateful for the privilege of observing the Universe from a place that is unique in both its astronomical quality and its cultural significance.
    We acknowledge with thanks the variable star observations from the AAVSO International Database contributed by observers worldwide and used in this research. 
\end{acknowledgements}

\bibliographystyle{aa}
\bibliography{biblio}

\begin{thebibliography}{101}
\expandafter\ifx\csname natexlab\endcsname\relax\def\natexlab#1{#1}\fi

\bibitem[{{Anderson} {et~al.}(2014){Anderson}, {Gonz{\'a}lez-Gait{\'a}n}, {Hamuy}, {Guti{\'e}rrez}, {Stritzinger}, {Olivares E.}, {Phillips}, {Schulze}, {Antezana}, {Bolt}, {Campillay}, {Castell{\'o}n}, {Contreras}, {de Jaeger}, {Folatelli}, {F{\"o}rster}, {Freedman}, {Gonz{\'a}lez}, {Hsiao}, {Krzemi{\'n}ski}, {Krisciunas}, {Maza}, {McCarthy}, {Morrell}, {Persson}, {Roth}, {Salgado}, {Suntzeff}, \& {Thomas-Osip}}]{Anderson2014}
{Anderson}, J.~P., {Gonz{\'a}lez-Gait{\'a}n}, S., {Hamuy}, M., {et~al.} 2014, \apj, 786, 67

\bibitem[{{Andrews} {et~al.}(2010){Andrews}, {Gallagher}, {Clayton}, {Sugerman}, {Chatelain}, {Clem}, {Welch}, {Barlow}, {Ercolano}, {Fabbri}, {Wesson}, \& {Meixner}}]{Andrews2010}
{Andrews}, J.~E., {Gallagher}, J.~S., {Clayton}, G.~C., {et~al.} 2010, \apj, 715, 541

\bibitem[{{Andrews} {et~al.}(2016){Andrews}, {Krafton}, {Clayton}, {Montiel}, {Wesson}, {Sugerman}, {Barlow}, {Matsuura}, \& {Drass}}]{Andrews2016}
{Andrews}, J.~E., {Krafton}, K.~M., {Clayton}, G.~C., {et~al.} 2016, \mnras, 457, 3241

\bibitem[{{Andrews} {et~al.}(2019){Andrews}, {Sand}, {Valenti}, {Smith}, {Dastidar}, {Sahu}, {Misra}, {Singh}, {Hiramatsu}, {Brown}, {Hosseinzadeh}, {Wyatt}, {Vinko}, {Anupama}, {Arcavi}, {Ashall}, {Benetti}, {Berton}, {Bostroem}, {Bulla}, {Burke}, {Chen}, {Chomiuk}, {Cikota}, {Congiu}, {Cseh}, {Davis}, {Elias-Rosa}, {Faran}, {Fraser}, {Galbany}, {Gall}, {Gal-Yam}, {Gangopadhyay}, {Gromadzki}, {Haislip}, {Howell}, {Hsiao}, {Inserra}, {Kankare}, {Kuncarayakti}, {Kouprianov}, {Kumar}, {Li}, {Lin}, {Maguire}, {Mazzali}, {McCully}, {Milne}, {Mo}, {Morrell}, {Nicholl}, {Ochner}, {Olivares}, {Pastorello}, {Patat}, {Phillips}, {Pignata}, {Prentice}, {Reguitti}, {Reichart}, {Rodr{\'\i}guez}, {Rui}, {Sanwal}, {S{\'a}rneczky}, {Shahbandeh}, {Singh}, {Smartt}, {Strader}, {Stritzinger}, {Szak{\'a}ts}, {Tartaglia}, {Wang}, {Wang}, {Wang}, {Wheeler}, {Xiang}, {Yaron}, {Young}, \& {Zhang}}]{Andrews2019}
{Andrews}, J.~E., {Sand}, D.~J., {Valenti}, S., {et~al.} 2019, \apj, 885, 43

\bibitem[{{Andrews} {et~al.}(2011){Andrews}, {Sugerman}, {Clayton}, {Gallagher}, {Barlow}, {Clem}, {Ercolano}, {Fabbri}, {Meixner}, {Otsuka}, {Welch}, \& {Wesson}}]{Andrews2011}
{Andrews}, J.~E., {Sugerman}, B.~E.~K., {Clayton}, G.~C., {et~al.} 2011, \apj, 731, 47

\bibitem[{{Beasor} {et~al.}(2020){Beasor}, {Davies}, {Smith}, {van Loon}, {Gehrz}, \& {Figer}}]{Beasor2020}
{Beasor}, E.~R., {Davies}, B., {Smith}, N., {et~al.} 2020, \mnras, 492, 5994

\bibitem[{{Bellm} {et~al.}(2019){Bellm}, {Kulkarni}, {Graham}, {Dekany}, {Smith}, {Riddle}, {Masci}, {Helou}, {Prince}, {Adams}, {Barbarino}, {Barlow}, {Bauer}, {Beck}, {Belicki}, {Biswas}, {Blagorodnova}, {Bodewits}, {Bolin}, {Brinnel}, {Brooke}, {Bue}, {Bulla}, {Burruss}, {Cenko}, {Chang}, {Connolly}, {Coughlin}, {Cromer}, {Cunningham}, {De}, {Delacroix}, {Desai}, {Duev}, {Eadie}, {Farnham}, {Feeney}, {Feindt}, {Flynn}, {Franckowiak}, {Frederick}, {Fremling}, {Gal-Yam}, {Gezari}, {Giomi}, {Goldstein}, {Golkhou}, {Goobar}, {Groom}, {Hacopians}, {Hale}, {Henning}, {Ho}, {Hover}, {Howell}, {Hung}, {Huppenkothen}, {Imel}, {Ip}, {Ivezi{\'c}}, {Jackson}, {Jones}, {Juric}, {Kasliwal}, {Kaspi}, {Kaye}, {Kelley}, {Kowalski}, {Kramer}, {Kupfer}, {Landry}, {Laher}, {Lee}, {Lin}, {Lin}, {Lunnan}, {Giomi}, {Mahabal}, {Mao}, {Miller}, {Monkewitz}, {Murphy}, {Ngeow}, {Nordin}, {Nugent}, {Ofek}, {Patterson}, {Penprase}, {Porter}, {Rauch}, {Rebbapragada}, {Reiley}, {Rigault}, {Rodriguez}, {van Roestel}, {Rusholme}, {van
  Santen}, {Schulze}, {Shupe}, {Singer}, {Soumagnac}, {Stein}, {Surace}, {Sollerman}, {Szkody}, {Taddia}, {Terek}, {Van Sistine}, {van Velzen}, {Vestrand}, {Walters}, {Ward}, {Ye}, {Yu}, {Yan}, \& {Zolkower}}]{Bellm2019}
{Bellm}, E.~C., {Kulkarni}, S.~R., {Graham}, M.~J., {et~al.} 2019, \pasp, 131, 018002

\bibitem[{{Berger} {et~al.}(2023){Berger}, {Keating}, {Margutti}, {Maeda}, {Alexander}, {Cendes}, {Eftekhari}, {Gurwell}, {Hiramatsu}, {Ho}, {Laskar}, {Rao}, \& {Williams}}]{berger2023}
{Berger}, E., {Keating}, G.~K., {Margutti}, R., {et~al.} 2023, \apjl, 951, L31

\bibitem[{{Bersten} {et~al.}(2024){Bersten}, {Orellana}, {Folatelli}, {Martinez}, {Piccirilli}, {Regna}, {Rom{\'a}n Aguilar}, \& {Ertini}}]{bersten2024}
{Bersten}, M.~C., {Orellana}, M., {Folatelli}, G., {et~al.} 2024, \aap, 681, L18

\bibitem[{{Black} {et~al.}(2017){Black}, {Milisavljevic}, {Margutti}, {Fesen}, {Patnaude}, \& {Parker}}]{Black2017}
{Black}, C.~S., {Milisavljevic}, D., {Margutti}, R., {et~al.} 2017, \apj, 848, 5

\bibitem[{{Bostroem} {et~al.}(2023){Bostroem}, {Pearson}, {Shrestha}, {Sand}, {Valenti}, {Jha}, {Andrews}, {Smith}, {Terreran}, {Green}, {Dong}, {Lundquist}, {Haislip}, {Hoang}, {Hosseinzadeh}, {Janzen}, {Jencson}, {Kouprianov}, {Paraskeva}, {Meza Retamal}, {Reichart}, {Arcavi}, {Bonanos}, {Coughlin}, {Dobson}, {Farah}, {Galbany}, {Guti{\'e}rrez}, {Hawley}, {Hebb}, {Hiramatsu}, {Howell}, {Iijima}, {Ilyin}, {Jhass}, {McCully}, {Moran}, {Morris}, {Mura}, {M{\"u}ller-Bravo}, {Munday}, {Newsome}, {Pabst}, {Ochner}, {Gonzalez}, {Pastorello}, {Pellegrino}, {Piscarreta}, {Ravi}, {Reguitti}, {Salo}, {Vink{\'o}}, {de Vos}, {Wheeler}, {Williams}, \& {Wyatt}}]{bostroem2023}
{Bostroem}, K.~A., {Pearson}, J., {Shrestha}, M., {et~al.} 2023, \apjl, 956, L5

\bibitem[{{Bostroem} {et~al.}(2024){Bostroem}, {Sand}, {Dessart}, {Smith}, {Jha}, {Valenti}, {Andrews}, {Dong}, {Filippenko}, {Gomez}, {Hiramatsu}, {Hoang}, {Hosseinzadeh}, {Howell}, {Jencson}, {Lundquist}, {McCully}, {Mehta}, {Meza-Retamal}, {Pearson}, {Ravi}, {Shrestha}, \& {Wyatt}}]{Bostroem2024}
{Bostroem}, K.~A., {Sand}, D.~J., {Dessart}, L., {et~al.} 2024, \apjl, 973, L47

\bibitem[{{Branch} {et~al.}(1981){Branch}, {Falk}, {McCall}, {Rybski}, {Uomoto}, \& {Wills}}]{Branch1981}
{Branch}, D., {Falk}, S.~W., {McCall}, M.~L., {et~al.} 1981, \apj, 244, 780

\bibitem[{{Bruch} {et~al.}(2023){Bruch}, {Gal-Yam}, {Yaron}, {Chen}, {Strotjohann}, {Irani}, {Zimmerman}, {Schulze}, {Yang}, {Kim}, {Bulla}, {Sollerman}, {Rigault}, {Ofek}, {Soumagnac}, {Masci}, {Fremling}, {Perley}, {Nordin}, {Cenko}, {Ho}, {Adams}, {Adreoni}, {Bellm}, {Blagorodnova}, {Burdge}, {De}, {Dekany}, {Dhawan}, {Drake}, {Duev}, {Graham}, {Graham}, {Jencson}, {Karamehmetoglu}, {Kasliwal}, {Kulkarni}, {Miller}, {Neill}, {Prince}, {Riddle}, {Rusholme}, {Sharma}, {Smith}, {Sravan}, {Taggart}, {Walters}, \& {Yan}}]{Bruch2023}
{Bruch}, R.~J., {Gal-Yam}, A., {Yaron}, O., {et~al.} 2023, \apj, 952, 119

\bibitem[{{Buta} \& {Keel}(2019)}]{Buta2019}
{Buta}, R.~J. \& {Keel}, W.~C. 2019, \mnras, 487, 832

\bibitem[{{Cardelli} {et~al.}(1989){Cardelli}, {Clayton}, \& {Mathis}}]{Cardelli1989}
{Cardelli}, J.~A., {Clayton}, G.~C., \& {Mathis}, J.~S. 1989, \apj, 345, 245

\bibitem[{{Chandra} {et~al.}(2024){Chandra}, {Chevalier}, {Maeda}, {Ray}, \& {Nayana}}]{Chandra2024}
{Chandra}, P., {Chevalier}, R.~A., {Maeda}, K., {Ray}, A.~K., \& {Nayana}, A.~J. 2024, \apjl, 963, L4

\bibitem[{{Chevalier} \& {Fransson}(2017)}]{Chevalier2017}
{Chevalier}, R.~A. \& {Fransson}, C. 2017, in Handbook of Supernovae, ed. A.~W. {Alsabti} \& P.~{Murdin}, 875

\bibitem[{{Chugai}(1992)}]{Chugai1992}
{Chugai}, N.~N. 1992, Soviet Astronomy Letters, 18, 239

\bibitem[{{Chugai} {et~al.}(2007){Chugai}, {Chevalier}, \& {Utrobin}}]{Chugai2007}
{Chugai}, N.~N., {Chevalier}, R.~A., \& {Utrobin}, V.~P. 2007, \apj, 662, 1136

\bibitem[{{Dessart} {et~al.}(2023){Dessart}, {Guti{\'e}rrez}, {Kuncarayakti}, {Fox}, \& {Filippenko}}]{Dessart2023}
{Dessart}, L., {Guti{\'e}rrez}, C.~P., {Kuncarayakti}, H., {Fox}, O.~D., \& {Filippenko}, A.~V. 2023, \aap, 675, A33

\bibitem[{{Dessart} \& {Hillier}(2022)}]{Dessart2022}
{Dessart}, L. \& {Hillier}, D.~J. 2022, \aap, 660, L9

\bibitem[{{Dessart} {et~al.}(2021){Dessart}, {Hillier}, {Sukhbold}, {Woosley}, \& {Janka}}]{Dessart2021}
{Dessart}, L., {Hillier}, D.~J., {Sukhbold}, T., {Woosley}, S.~E., \& {Janka}, H.~T. 2021, \aap, 652, A64

\bibitem[{{Dhungana} {et~al.}(2016){Dhungana}, {Kehoe}, {Vinko}, {Silverman}, {Wheeler}, {Zheng}, {Marion}, {Fox}, {Akerlof}, {Biro}, {Borkovits}, {Cenko}, {Clubb}, {Filippenko}, {Ferrante}, {Gibson}, {Graham}, {Hegedus}, {Kelly}, {Kelemen}, {Lee}, {Marschalko}, {Moln{\'a}r}, {Nagy}, {Ordasi}, {Pal}, {Sarneczky}, {Shivvers}, {Szakats}, {Szalai}, {Szegedi-Elek}, {Sz{\'e}kely}, {Szing}, {Tak{\'a}ts}, \& {Vida}}]{Dhungana2016}
{Dhungana}, G., {Kehoe}, R., {Vinko}, J., {et~al.} 2016, \apj, 822, 6

\bibitem[{{Dong} {et~al.}(2023){Dong}, {Sand}, {Valenti}, {Bostroem}, {Andrews}, {Hosseinzadeh}, {Hoang}, {Janzen}, {Jencson}, {Lundquist}, {Meza Retamal}, {Pearson}, {Shrestha}, {Haislip}, {Kouprianov}, \& {Reichart}}]{Dong2023}
{Dong}, Y., {Sand}, D.~J., {Valenti}, S., {et~al.} 2023, \apj, 957, 28

\bibitem[{{Elmhamdi}(2011)}]{Elmhamdi2011}
{Elmhamdi}, A. 2011, \actaa, 61, 179

\bibitem[{{Elmhamdi} {et~al.}(2004){Elmhamdi}, {Danziger}, {Cappellaro}, {Della Valle}, {Gouiffes}, {Phillips}, \& {Turatto}}]{Elmhamdi2004}
{Elmhamdi}, A., {Danziger}, I.~J., {Cappellaro}, E., {et~al.} 2004, \aap, 426, 963

\bibitem[{{Fang} {et~al.}(2025){Fang}, {Moriya}, {Ferrari}, {Maeda}, {Folatelli}, {Ertini}, {Kuncarayakti}, {Andrews}, \& {Matsumoto}}]{Fang2025}
{Fang}, Q., {Moriya}, T.~J., {Ferrari}, L., {et~al.} 2025, \apj, 978, 36

\bibitem[{{Faran} {et~al.}(2014){Faran}, {Poznanski}, {Filippenko}, {Chornock}, {Foley}, {Ganeshalingam}, {Leonard}, {Li}, {Modjaz}, {Nakar}, {Serduke}, \& {Silverman}}]{Faran2014}
{Faran}, T., {Poznanski}, D., {Filippenko}, A.~V., {et~al.} 2014, \mnras, 442, 844

\bibitem[{{Ferrari} {et~al.}(2024){Ferrari}, {Folatelli}, {Ertini}, {Kuncarayakti}, \& {Andrews}}]{ferrari2024}
{Ferrari}, L., {Folatelli}, G., {Ertini}, K., {Kuncarayakti}, H., \& {Andrews}, J.~E. 2024, \aap, 687, L20

\bibitem[{{Filippenko} {et~al.}(1994){Filippenko}, {Matheson}, \& {Barth}}]{Filippenko1994}
{Filippenko}, A.~V., {Matheson}, T., \& {Barth}, A.~J. 1994, \aj, 108, 2220

\bibitem[{{F{\"o}rster} {et~al.}(2021){F{\"o}rster}, {Cabrera-Vives}, {Castillo-Navarrete}, {Est{\'e}vez}, {S{\'a}nchez-S{\'a}ez}, {Arredondo}, {Bauer}, {Carrasco-Davis}, {Catelan}, {Elorrieta}, {Eyheramendy}, {Huijse}, {Pignata}, {Reyes}, {Reyes}, {Rodr{\'\i}guez-Mancini}, {Ruz-Mieres}, {Valenzuela}, {{\'A}lvarez-Maldonado}, {Astorga}, {Borissova}, {Clocchiatti}, {De Cicco}, {Donoso-Oliva}, {Hern{\'a}ndez-Garc{\'\i}a}, {Graham}, {Jord{\'a}n}, {Kurtev}, {Mahabal}, {Maureira}, {Mu{\~n}oz-Arancibia}, {Molina-Ferreiro}, {Moya}, {Palma}, {P{\'e}rez-Carrasco}, {Protopapas}, {Romero}, {Sabatini-Gacitua}, {S{\'a}nchez}, {San Mart{\'\i}n}, {Sep{\'u}lveda-Cobo}, {Vera}, \& {Vergara}}]{Forster2021}
{F{\"o}rster}, F., {Cabrera-Vives}, G., {Castillo-Navarrete}, E., {et~al.} 2021, \aj, 161, 242

\bibitem[{{F{\"o}rster} {et~al.}(2018){F{\"o}rster}, {Moriya}, {Maureira}, {Anderson}, {Blinnikov}, {Bufano}, {Cabrera-Vives}, {Clocchiatti}, {de Jaeger}, {Est{\'e}vez}, {Galbany}, {Gonz{\'a}lez-Gait{\'a}n}, {Gr{\"a}fener}, {Hamuy}, {Hsiao}, {Huentelemu}, {Huijse}, {Kuncarayakti}, {Mart{\'\i}nez}, {Medina}, {Olivares E.}, {Pignata}, {Razza}, {Reyes}, {San Mart{\'\i}n}, {Smith}, {Vera}, {Vivas}, {de Ugarte Postigo}, {Yoon}, {Ashall}, {Fraser}, {Gal-Yam}, {Kankare}, {Le Guillou}, {Mazzali}, {Walton}, \& {Young}}]{Forster2018}
{F{\"o}rster}, F., {Moriya}, T.~J., {Maureira}, J.~C., {et~al.} 2018, Nature Astronomy, 2, 808

\bibitem[{{Fransson} \& {Chevalier}(1987)}]{Fransson1987}
{Fransson}, C. \& {Chevalier}, R.~A. 1987, \apjl, 322, L15

\bibitem[{{Fransson} \& {Chevalier}(1989)}]{Fransson1989}
{Fransson}, C. \& {Chevalier}, R.~A. 1989, \apj, 343, 323

\bibitem[{{Grefenstette} {et~al.}(2023){Grefenstette}, {Brightman}, {Earnshaw}, {Harrison}, \& {Margutti}}]{grefenstette2023}
{Grefenstette}, B.~W., {Brightman}, M., {Earnshaw}, H.~P., {Harrison}, F.~A., \& {Margutti}, R. 2023, \apjl, 952, L3

\bibitem[{{Guti{\'e}rrez} {et~al.}(2017){Guti{\'e}rrez}, {Anderson}, {Hamuy}, {Morrell}, {Gonz{\'a}lez-Gaitan}, {Stritzinger}, {Phillips}, {Galbany}, {Folatelli}, {Dessart}, {Contreras}, {Della Valle}, {Freedman}, {Hsiao}, {Krisciunas}, {Madore}, {Maza}, {Suntzeff}, {Prieto}, {Gonz{\'a}lez}, {Cappellaro}, {Navarrete}, {Pizzella}, {Ruiz}, {Smith}, \& {Turatto}}]{Gutierrez2017}
{Guti{\'e}rrez}, C.~P., {Anderson}, J.~P., {Hamuy}, M., {et~al.} 2017, \apj, 850, 89

\bibitem[{{Guti{\'e}rrez} {et~al.}(2020){Guti{\'e}rrez}, {Pastorello}, {Jerkstrand}, {Galbany}, {Sullivan}, {Anderson}, {Taubenberger}, {Kuncarayakti}, {Gonz{\'a}lez-Gait{\'a}n}, {Wiseman}, {Inserra}, {Fraser}, {Maguire}, {Smartt}, {M{\"u}ller-Bravo}, {Arcavi}, {Benetti}, {Bersier}, {Bose}, {Bostroem}, {Burke}, {Chen}, {Chen}, {Della Valle}, {Dong}, {Gal-Yam}, {Gromadzki}, {Hiramatsu}, {Holoien}, {Hosseinzadeh}, {Howell}, {Kankare}, {Kochanek}, {McCully}, {Nicholl}, {Pignata}, {Prieto}, {Shappee}, {Taggart}, {Tomasella}, {Valenti}, \& {Young}}]{Gutierrez2020}
{Guti{\'e}rrez}, C.~P., {Pastorello}, A., {Jerkstrand}, A., {et~al.} 2020, \mnras, 499, 974

\bibitem[{{Hiramatsu} {et~al.}(2023){Hiramatsu}, {Tsuna}, {Berger}, {Itagaki}, {Goldberg}, {Gomez}, {Kishalay}, {Hosseinzadeh}, {Bostroem}, {Brown}, {Arcavi}, {Bieryla}, {Blanchard}, {Esquerdo}, {Farah}, {Howell}, {Matsumoto}, {McCully}, {Newsome}, {Gonzalez}, {Pellegrino}, {Rhee}, {Terreran}, {Vink{\'o}}, \& {Wheeler}}]{hiramatsu2023}
{Hiramatsu}, D., {Tsuna}, D., {Berger}, E., {et~al.} 2023, \apjl, 955, L8

\bibitem[{{Hook} {et~al.}(2004){Hook}, {J{\o}rgensen}, {Allington-Smith}, {Davies}, {Metcalfe}, {Murowinski}, \& {Crampton}}]{hook2004}
{Hook}, I.~M., {J{\o}rgensen}, I., {Allington-Smith}, J.~R., {et~al.} 2004, \pasp, 116, 425

\bibitem[{{Hosseinzadeh} {et~al.}(2023){Hosseinzadeh}, {Farah}, {Shrestha}, {Sand}, {Dong}, {Brown}, {Bostroem}, {Valenti}, {Jha}, {Andrews}, {Arcavi}, {Haislip}, {Hiramatsu}, {Hoang}, {Howell}, {Janzen}, {Jencson}, {Kouprianov}, {Lundquist}, {McCully}, {Meza Retamal}, {Modjaz}, {Newsome}, {Padilla Gonzalez}, {Pearson}, {Pellegrino}, {Ravi}, {Reichart}, {Smith}, {Terreran}, \& {Vink{\'o}}}]{hosseinzadeh2023}
{Hosseinzadeh}, G., {Farah}, J., {Shrestha}, M., {et~al.} 2023, \apjl, 953, L16

\bibitem[{{Inserra} {et~al.}(2011){Inserra}, {Turatto}, {Pastorello}, {Benetti}, {Cappellaro}, {Pumo}, {Zampieri}, {Agnoletto}, {Bufano}, {Botticella}, {Della Valle}, {Elias Rosa}, {Iijima}, {Spiro}, \& {Valenti}}]{Inserra2011}
{Inserra}, C., {Turatto}, M., {Pastorello}, A., {et~al.} 2011, \mnras, 417, 261

\bibitem[{{Iwata} {et~al.}(2025){Iwata}, {Akimoto}, {Matsuoka}, {Maeda}, {Yonekura}, {Tominaga}, {Moriya}, {Fujisawa}, {Niinuma}, {Yoon}, {Lee}, {Jung}, \& {Byun}}]{Iwata2025}
{Iwata}, Y., {Akimoto}, M., {Matsuoka}, T., {et~al.} 2025, \apj, 978, 138

\bibitem[{{Jacobson-Gal{\'a}n} {et~al.}(2024){Jacobson-Gal{\'a}n}, {Dessart}, {Davis}, {Kilpatrick}, {Margutti}, {Foley}, {Chornock}, {Terreran}, {Hiramatsu}, {Newsome}, {Padilla Gonzalez}, {Pellegrino}, {Howell}, {Filippenko}, {Anderson}, {Angus}, {Auchettl}, {Bostroem}, {Brink}, {Cartier}, {Coulter}, {de Boer}, {Drout}, {Earl}, {Ertini}, {Farah}, {Farias}, {Gall}, {Gao}, {Gerlach}, {Guo}, {Haynie}, {Hosseinzadeh}, {Ibik}, {Jha}, {Jones}, {Langeroodi}, {LeBaron}, {Magnier}, {Piro}, {Raimundo}, {Rest}, {Rest}, {Rich}, {Rojas-Bravo}, {Sears}, {Taggart}, {Villar}, {Wainscoat}, {Wang}, {Wasserman}, {Yan}, {Yang}, {Zhang}, \& {Zheng}}]{Jacobson-Galan2024}
{Jacobson-Gal{\'a}n}, W.~V., {Dessart}, L., {Davis}, K.~W., {et~al.} 2024, \apj, 970, 189

\bibitem[{{Jacobson-Gal{\'a}n} {et~al.}(2023){Jacobson-Gal{\'a}n}, {Dessart}, {Margutti}, {Chornock}, {Foley}, {Kilpatrick}, {Jones}, {Taggart}, {Angus}, {Bhattacharjee}, {Braff}, {Brethauer}, {Burgasser}, {Cao}, {Carlile}, {Chambers}, {Coulter}, {Dominguez-Ruiz}, {Dickinson}, {de Boer}, {Gagliano}, {Gall}, {Gao}, {Gates}, {Gomez}, {Guolo}, {Halford}, {Hjorth}, {Huber}, {Johnson}, {Karpoor}, {Laskar}, {LeBaron}, {Li}, {Lin}, {Loch}, {Lynam}, {Magnier}, {Maloney}, {Matthews}, {McDonald}, {Miao}, {Milisavljevic}, {Pan}, {Pradyumna}, {Ransome}, {Rees}, {Rest}, {Rojas-Bravo}, {Sandford}, {Ascencio}, {Sanjaripour}, {Savino}, {Sears}, {Sharei}, {Smartt}, {Softich}, {Theissen}, {Tinyanont}, {Tohfa}, {Villar}, {Wang}, {Wainscoat}, {Westerling}, {Wiston}, {Wozniak}, {Yadavalli}, \& {Zenati}}]{jacobsongalan2023}
{Jacobson-Gal{\'a}n}, W.~V., {Dessart}, L., {Margutti}, R., {et~al.} 2023, \apjl, 954, L42

\bibitem[{{Jencson} {et~al.}(2023){Jencson}, {Pearson}, {Beasor}, {Lau}, {Andrews}, {Bostroem}, {Dong}, {Engesser}, {Gomez}, {Guolo}, {Hoang}, {Hosseinzadeh}, {Jha}, {Karambelkar}, {Kasliwal}, {Lundquist}, {Meza Retamal}, {Rest}, {Sand}, {Shahbandeh}, {Shrestha}, {Smith}, {Strader}, {Valenti}, {Wang}, \& {Zenati}}]{jencson2023}
{Jencson}, J.~E., {Pearson}, J., {Beasor}, E.~R., {et~al.} 2023, \apjl, 952, L30

\bibitem[{{Jerkstrand} {et~al.}(2018){Jerkstrand}, {Ertl}, {Janka}, {M{\"u}ller}, {Sukhbold}, \& {Woosley}}]{Jerkstrand2018}
{Jerkstrand}, A., {Ertl}, T., {Janka}, H.~T., {et~al.} 2018, \mnras, 475, 277

\bibitem[{{Jerkstrand} {et~al.}(2012){Jerkstrand}, {Fransson}, {Maguire}, {Smartt}, {Ergon}, \& {Spyromilio}}]{Jerkstrand2012}
{Jerkstrand}, A., {Fransson}, C., {Maguire}, K., {et~al.} 2012, \aap, 546, A28

\bibitem[{{Jerkstrand} {et~al.}(2014){Jerkstrand}, {Smartt}, {Fraser}, {Fransson}, {Sollerman}, {Taddia}, \& {Kotak}}]{Jerkstrand2014}
{Jerkstrand}, A., {Smartt}, S.~J., {Fraser}, M., {et~al.} 2014, \mnras, 439, 3694

\bibitem[{{Kilpatrick} {et~al.}(2023){Kilpatrick}, {Foley}, {Jacobson-Gal{\'a}n}, {Piro}, {Smartt}, {Drout}, {Gagliano}, {Gall}, {Hjorth}, {Jones}, {Mandel}, {Margutti}, {Ramirez-Ruiz}, {Ransome}, {Villar}, {Coulter}, {Gao}, {Matthews}, {Taggart}, \& {Zenati}}]{kilpatrick2023}
{Kilpatrick}, C.~D., {Foley}, R.~J., {Jacobson-Gal{\'a}n}, W.~V., {et~al.} 2023, \apjl, 952, L23

\bibitem[{{Kotak} {et~al.}(2009){Kotak}, {Meikle}, {Farrah}, {Gerardy}, {Foley}, {Van Dyk}, {Fransson}, {Lundqvist}, {Sollerman}, {Fesen}, {Filippenko}, {Mattila}, {Silverman}, {Andersen}, {H{\"o}flich}, {Pozzo}, \& {Wheeler}}]{Kotak2009}
{Kotak}, R., {Meikle}, W.~P.~S., {Farrah}, D., {et~al.} 2009, \apj, 704, 306

\bibitem[{{Kumar} {et~al.}(2025){Kumar}, {Dastidar}, {Maund}, {Singleton}, \& {Sun}}]{Kumar2025}
{Kumar}, A., {Dastidar}, R., {Maund}, J.~R., {Singleton}, A.~J., \& {Sun}, N.-C. 2025, \mnras, 538, 659

\bibitem[{{Kuncarayakti} {et~al.}(2015){Kuncarayakti}, {Maeda}, {Bersten}, {Folatelli}, {Morrell}, {Hsiao}, {Gonz{\'a}lez-Gait{\'a}n}, {Anderson}, {Hamuy}, {de Jaeger}, {Guti{\'e}rrez}, \& {Kawabata}}]{Kuncarayakti2015}
{Kuncarayakti}, H., {Maeda}, K., {Bersten}, M.~C., {et~al.} 2015, \aap, 579, A95

\bibitem[{{Labrie} {et~al.}(2023){Labrie}, {Simpson}, {Cardenes}, {Turner}, {Soraisam}, {Quint}, {Oberdorf}, {Placco}, {Berke}, {Smirnova}, {Conseil}, {Vacca}, \& {Thomas-Osip}}]{Labrie2023}
{Labrie}, K., {Simpson}, C., {Cardenes}, R., {et~al.} 2023, Research Notes of the American Astronomical Society, 7, 214

\bibitem[{{Li} {et~al.}(2024){Li}, {Hu}, {Li}, {Yang}, {Wang}, {Yan}, {Hu}, {Zhang}, {Mao}, {Riise}, {Gao}, {Sun}, {Liu}, {Xiong}, {Wang}, {Mo}, {Iskandar}, {Xi}, {Xiang}, {Wang}, {Sun}, {Zhang}, {Chen}, {Lin}, {Guo}, {Liu}, {Cai}, {Zhou}, {Zhao}, {Chen}, {Zheng}, {Li}, {Zhang}, {Xu}, {Lyu}, {Castro-Tirado}, {Chufarin}, {Potapov}, {Ionov}, {Korotkiy}, {Nazarov}, {Sokolovsky}, {Hamann}, \& {Herman}}]{Li2024}
{Li}, G., {Hu}, M., {Li}, W., {et~al.} 2024, \nat, 627, 754

\bibitem[{{Li} {et~al.}(2025){Li}, {Wang}, {Yang}, {Pastorello}, {Reguitti}, {Valerin}, {Ochner}, {Cai}, {Iijima}, {Munari}, {Salmaso}, {Farina}, {Cazzola}, {Trabacchin}, {Fiscale}, {Ciroi}, {Mura}, {Siviero}, {Cabras}, {Pabst}, {Taubenberger}, {Vogl}, {Fiorin}, {Liu}, {Chen}, {Xiang}, {Mo}, {Li}, {Wang}, {Zhang}, {Zhai}, {Mirzaqulov}, {Ehgamberdiev}, {Filippenko}, {Yan}, {Hu}, {Ma}, {Xia}, {Gao}, \& {Li}}]{Li2025arXiv250403856L}
{Li}, G., {Wang}, X., {Yang}, Y., {et~al.} 2025, arXiv e-prints, arXiv:2504.03856

\bibitem[{{Liu} {et~al.}(2023){Liu}, {Chen}, {Er}, {Zeimann}, {Vink{\'o}}, {Wheeler}, {Cooper}, {Davis}, {Farrow}, {Gebhardt}, {Guo}, {Hill}, {House}, {Kollatschny}, {Kong}, {Kumar}, {Liu}, {Tuttle}, {Endl}, {Duke}, {Cochran}, {Zhang}, \& {Liu}}]{liu2023}
{Liu}, C., {Chen}, X., {Er}, X., {et~al.} 2023, \apjl, 958, L37

\bibitem[{{Lundquist} {et~al.}(2023){Lundquist}, {O'Meara}, \& {Walawender}}]{Lundquist2023}
{Lundquist}, M., {O'Meara}, J., \& {Walawender}, J. 2023, Transient Name Server AstroNote, 160, 1

\bibitem[{{Maeda} {et~al.}(2015){Maeda}, {Hattori}, {Milisavljevic}, {Folatelli}, {Drout}, {Kuncarayakti}, {Margutti}, {Kamble}, {Soderberg}, {Tanaka}, {Kawabata}, {Kawabata}, {Yamanaka}, {Nomoto}, {Kim}, {Simon}, {Phillips}, {Parrent}, {Nakaoka}, {Moriya}, {Suzuki}, {Takaki}, {Ishigaki}, {Sakon}, {Tajitsu}, \& {Iye}}]{Maeda2015}
{Maeda}, K., {Hattori}, T., {Milisavljevic}, D., {et~al.} 2015, \apj, 807, 35

\bibitem[{{Maguire} {et~al.}(2010){Maguire}, {Di Carlo}, {Smartt}, {Pastorello}, {Tsvetkov}, {Benetti}, {Spiro}, {Arkharov}, {Beccari}, {Botticella}, {Cappellaro}, {Cristallo}, {Dolci}, {Elias-Rosa}, {Fiaschi}, {Gorshanov}, {Harutyunyan}, {Larionov}, {Navasardyan}, {Pietrinferni}, {Raimondo}, {di Rico}, {Valenti}, {Valentini}, \& {Zampieri}}]{Maguire2010}
{Maguire}, K., {Di Carlo}, E., {Smartt}, S.~J., {et~al.} 2010, \mnras, 404, 981

\bibitem[{{Maguire} {et~al.}(2012){Maguire}, {Jerkstrand}, {Smartt}, {Fransson}, {Pastorello}, {Benetti}, {Valenti}, {Bufano}, \& {Leloudas}}]{Maguire2012}
{Maguire}, K., {Jerkstrand}, A., {Smartt}, S.~J., {et~al.} 2012, \mnras, 420, 3451

\bibitem[{{Martinez} {et~al.}(2024){Martinez}, {Bersten}, {Folatelli}, {Orellana}, \& {Ertini}}]{Martinez2024}
{Martinez}, L., {Bersten}, M.~C., {Folatelli}, G., {Orellana}, M., \& {Ertini}, K. 2024, \aap, 683, A154

\bibitem[{{Masci} {et~al.}(2019){Masci}, {Laher}, {Rusholme}, {Shupe}, {Groom}, {Surace}, {Jackson}, {Monkewitz}, {Beck}, {Flynn}, {Terek}, {Landry}, {Hacopians}, {Desai}, {Howell}, {Brooke}, {Imel}, {Wachter}, {Ye}, {Lin}, {Cenko}, {Cunningham}, {Rebbapragada}, {Bue}, {Miller}, {Mahabal}, {Bellm}, {Patterson}, {Juri{\'c}}, {Golkhou}, {Ofek}, {Walters}, {Graham}, {Kasliwal}, {Dekany}, {Kupfer}, {Burdge}, {Cannella}, {Barlow}, {Van Sistine}, {Giomi}, {Fremling}, {Blagorodnova}, {Levitan}, {Riddle}, {Smith}, {Helou}, {Prince}, \& {Kulkarni}}]{Masci2019}
{Masci}, F.~J., {Laher}, R.~R., {Rusholme}, B., {et~al.} 2019, \pasp, 131, 018003

\bibitem[{{Matheson} {et~al.}(2000){Matheson}, {Filippenko}, {Ho}, {Barth}, \& {Leonard}}]{Matheson2000}
{Matheson}, T., {Filippenko}, A.~V., {Ho}, L.~C., {Barth}, A.~J., \& {Leonard}, D.~C. 2000, \aj, 120, 1499

\bibitem[{{Matsuoka} {et~al.}(2019){Matsuoka}, {Maeda}, {Lee}, \& {Yasuda}}]{Matsuoka2019}
{Matsuoka}, T., {Maeda}, K., {Lee}, S.-H., \& {Yasuda}, H. 2019, \apj, 885, 41

\bibitem[{{Mauerhan} {et~al.}(2017){Mauerhan}, {Van Dyk}, {Johansson}, {Hu}, {Fox}, {Wang}, {Graham}, {Filippenko}, \& {Shivvers}}]{Mauerhan2017}
{Mauerhan}, J.~C., {Van Dyk}, S.~D., {Johansson}, J., {et~al.} 2017, \apj, 834, 118

\bibitem[{{Michel} {et~al.}(2025){Michel}, {Mazzali}, {Perley}, {Hinds}, \& {Wise}}]{Michel2025}
{Michel}, P.~D., {Mazzali}, P.~A., {Perley}, D.~A., {Hinds}, K.~R., \& {Wise}, J.~L. 2025, \mnras

\bibitem[{{Moriya}(2021)}]{Moriya2021}
{Moriya}, T.~J. 2021, \mnras, 503, L28

\bibitem[{{Moriya} {et~al.}(2018){Moriya}, {F{\"o}rster}, {Yoon}, {Gr{\"a}fener}, \& {Blinnikov}}]{Moriya2018}
{Moriya}, T.~J., {F{\"o}rster}, F., {Yoon}, S.-C., {Gr{\"a}fener}, G., \& {Blinnikov}, S.~I. 2018, \mnras, 476, 2840

\bibitem[{{Moriya} \& {Singh}(2024)}]{Moriya2024}
{Moriya}, T.~J. \& {Singh}, A. 2024, \pasj, 76, 1050

\bibitem[{{Moriya} {et~al.}(2023){Moriya}, {Subrayan}, {Milisavljevic}, \& {Blinnikov}}]{Moriya2023}
{Moriya}, T.~J., {Subrayan}, B.~M., {Milisavljevic}, D., \& {Blinnikov}, S.~I. 2023, \pasj, 75, 634

\bibitem[{{Moriya} {et~al.}(2017){Moriya}, {Yoon}, {Gr{\"a}fener}, \& {Blinnikov}}]{Moriya2017}
{Moriya}, T.~J., {Yoon}, S.-C., {Gr{\"a}fener}, G., \& {Blinnikov}, S.~I. 2017, \mnras, 469, L108

\bibitem[{{Neustadt} {et~al.}(2024){Neustadt}, {Kochanek}, \& {Smith}}]{neustadt2024}
{Neustadt}, J.~M.~M., {Kochanek}, C.~S., \& {Smith}, M.~R. 2024, \mnras, 527, 5366

\bibitem[{{Niu} {et~al.}(2023){Niu}, {Sun}, {Maund}, {Zhang}, {Zhao}, \& {Liu}}]{niu2023}
{Niu}, Z., {Sun}, N.-C., {Maund}, J.~R., {et~al.} 2023, \apjl, 955, L15

\bibitem[{{Patat} {et~al.}(1995){Patat}, {Chugai}, \& {Mazzali}}]{Patat1995}
{Patat}, F., {Chugai}, N., \& {Mazzali}, P.~A. 1995, \aap, 299, 715

\bibitem[{{Pledger} \& {Shara}(2023)}]{pledger2023}
{Pledger}, J.~L. \& {Shara}, M.~M. 2023, \apjl, 953, L14

\bibitem[{{Prieto} {et~al.}(2012){Prieto}, {Lee}, {Drake}, {McNaught}, {Garradd}, {Beacom}, {Beshore}, {Catelan}, {Djorgovski}, {Pojmanski}, {Stanek}, \& {Szczygie{\l}}}]{Prieto2012}
{Prieto}, J.~L., {Lee}, J.~C., {Drake}, A.~J., {et~al.} 2012, \apj, 745, 70

\bibitem[{{Qin} {et~al.}(2024){Qin}, {Zhang}, {Bloom}, {Sollerman}, {Zimmerman}, {Irani}, {Schulze}, {Gal-Yam}, {Kasliwal}, {Coughlin}, {Perley}, {Fremling}, \& {Kulkarni}}]{Qin2024}
{Qin}, Y.-J., {Zhang}, K., {Bloom}, J., {et~al.} 2024, \mnras, 534, 271

\bibitem[{{Ransome} {et~al.}(2024){Ransome}, {Villar}, {Tartaglia}, {Gonzalez}, {Jacobson-Gal{\'a}n}, {Kilpatrick}, {Margutti}, {Foley}, {Grayling}, {Ni}, {Yarza}, {Ye}, {Auchettl}, {de Boer}, {Chambers}, {Coulter}, {Drout}, {Farias}, {Gall}, {Gao}, {Huber}, {Ibik}, {Jones}, {Khetan}, {Lin}, {Politsch}, {Raimundo}, {Rest}, {Wainscoat}, {Yadavalli}, \& {Zenati}}]{Ransome2024}
{Ransome}, C.~L., {Villar}, V.~A., {Tartaglia}, A., {et~al.} 2024, \apj, 965, 93

\bibitem[{{Riess} {et~al.}(2022){Riess}, {Yuan}, {Macri}, {Scolnic}, {Brout}, {Casertano}, {Jones}, {Murakami}, {Anand}, {Breuval}, {Brink}, {Filippenko}, {Hoffmann}, {Jha}, {D'arcy Kenworthy}, {Mackenty}, {Stahl}, \& {Zheng}}]{Riess2022}
{Riess}, A.~G., {Yuan}, W., {Macri}, L.~M., {et~al.} 2022, \apjl, 934, L7

\bibitem[{{Schlafly} \& {Finkbeiner}(2011)}]{Schlafly2011}
{Schlafly}, E.~F. \& {Finkbeiner}, D.~P. 2011, \apj, 737, 103

\bibitem[{{Silverman} {et~al.}(2017){Silverman}, {Pickett}, {Wheeler}, {Filippenko}, {Vink{\'o}}, {Marion}, {Cenko}, {Chornock}, {Clubb}, {Foley}, {Graham}, {Kelly}, {Matheson}, \& {Shields}}]{Silverman2017}
{Silverman}, J.~M., {Pickett}, S., {Wheeler}, J.~C., {et~al.} 2017, \mnras, 467, 369

\bibitem[{{Singh} {et~al.}(2024){Singh}, {Teja}, {Moriya}, {Maeda}, {Kawabata}, {Tanaka}, {Imazawa}, {Nakaoka}, {Gangopadhyay}, {Yamanaka}, {Swain}, {Sahu}, {Anupama}, {Kumar}, {Anche}, {Sano}, {Raj}, {Agnihotri}, {Bhalerao}, {Bisht}, {Bisht}, {Belwal}, {Chakrabarti}, {Fujii}, {Nagayama}, {Matsumoto}, {Hamada}, {Kawabata}, {Kumar}, {Kumar}, {Malkan}, {Smith}, {Sakagami}, {Taguchi}, {Tominaga}, \& {Watanabe}}]{Singh2024}
{Singh}, A., {Teja}, R.~S., {Moriya}, T.~J., {et~al.} 2024, \apj, 975, 132

\bibitem[{{Smartt}(2015)}]{smartt2015}
{Smartt}, S.~J. 2015, \pasa, 32, e016

\bibitem[{{Smartt} {et~al.}(2009){Smartt}, {Eldridge}, {Crockett}, \& {Maund}}]{Smartt2009}
{Smartt}, S.~J., {Eldridge}, J.~J., {Crockett}, R.~M., \& {Maund}, J.~R. 2009, \mnras, 395, 1409

\bibitem[{{Smith}(2014)}]{Smith2014}
{Smith}, N. 2014, \araa, 52, 487

\bibitem[{{Soraisam} {et~al.}(2023){Soraisam}, {Szalai}, {Van Dyk}, {Andrews}, {Srinivasan}, {Chun}, {Matheson}, {Scicluna}, \& {Vasquez-Torres}}]{soraisam2023}
{Soraisam}, M.~D., {Szalai}, T., {Van Dyk}, S.~D., {et~al.} 2023, \apj, 957, 64

\bibitem[{{Tartaglia} {et~al.}(2021){Tartaglia}, {Sand}, {Groh}, {Valenti}, {Wyatt}, {Bostroem}, {Brown}, {Yang}, {Burke}, {Chen}, {Davis}, {F{\"o}rster}, {Galbany}, {Haislip}, {Hiramatsu}, {Hosseinzadeh}, {Howell}, {Hsiao}, {Jha}, {Kouprianov}, {Kuncarayakti}, {Lyman}, {McCully}, {Phillips}, {Rau}, {Reichart}, {Shahbandeh}, \& {Strader}}]{Tartaglia2021}
{Tartaglia}, L., {Sand}, D.~J., {Groh}, J.~H., {et~al.} 2021, \apj, 907, 52

\bibitem[{{Teja} {et~al.}(2023){Teja}, {Singh}, {Basu}, {Anupama}, {Sahu}, {Dutta}, {Swain}, {Nakaoka}, {Pathak}, {Bhalerao}, {Barway}, {Kumar}, {A.~J.}, {Imazawa}, {Kumar}, \& {Kawabata}}]{teja2023}
{Teja}, R.~S., {Singh}, A., {Basu}, J., {et~al.} 2023, \apjl, 954, L12

\bibitem[{{Terreran} {et~al.}(2016){Terreran}, {Jerkstrand}, {Benetti}, {Smartt}, {Ochner}, {Tomasella}, {Howell}, {Morales-Garoffolo}, {Harutyunyan}, {Kankare}, {Arcavi}, {Cappellaro}, {Elias-Rosa}, {Hosseinzadeh}, {Kangas}, {Pastorello}, {Tartaglia}, {Turatto}, {Valenti}, {Wiggins}, \& {Yuan}}]{Terreran2016}
{Terreran}, G., {Jerkstrand}, A., {Benetti}, S., {et~al.} 2016, \mnras, 462, 137

\bibitem[{{Uomoto}(1986)}]{Uomoto1986}
{Uomoto}, A. 1986, \apjl, 310, L35

\bibitem[{{Van Dyk} {et~al.}(2012){Van Dyk}, {Cenko}, {Poznanski}, {Arcavi}, {Gal-Yam}, {Filippenko}, {Silverio}, {Stockton}, {Cuillandre}, {Marcy}, {Howard}, \& {Isaacson}}]{vandyk2012}
{Van Dyk}, S.~D., {Cenko}, S.~B., {Poznanski}, D., {et~al.} 2012, \apj, 756, 131

\bibitem[{{Van Dyk} {et~al.}(2024){Van Dyk}, {Srinivasan}, {Andrews}, {Soraisam}, {Szalai}, {Howell}, {Isaacson}, {Matheson}, {Petigura}, {Scicluna}, {Stephens}, {Van Zandt}, {Zheng}, {Chun}, \& {Fillippenko}}]{vandyk2024}
{Van Dyk}, S.~D., {Srinivasan}, S., {Andrews}, J.~E., {et~al.} 2024, \apj, 968, 27

\bibitem[{{Vink} {et~al.}(2001){Vink}, {de Koter}, \& {Lamers}}]{Vink2001}
{Vink}, J.~S., {de Koter}, A., \& {Lamers}, H.~J.~G.~L.~M. 2001, \aap, 369, 574

\bibitem[{{Weil} {et~al.}(2020){Weil}, {Fesen}, {Patnaude}, \& {Milisavljevic}}]{Weil2020}
{Weil}, K.~E., {Fesen}, R.~A., {Patnaude}, D.~J., \& {Milisavljevic}, D. 2020, \apj, 900, 11

\bibitem[{{Xiang} {et~al.}(2024){Xiang}, {Mo}, {Wang}, {Wang}, {Zhang}, {Lin}, \& {Wang}}]{Xiang2024}
{Xiang}, D., {Mo}, J., {Wang}, L., {et~al.} 2024, Science China Physics, Mechanics, and Astronomy, 67, 219514

\bibitem[{{Yaron} \& {Gal-Yam}(2012)}]{yaron2012}
{Yaron}, O. \& {Gal-Yam}, A. 2012, \pasp, 124, 668

\bibitem[{{Yaron} {et~al.}(2017){Yaron}, {Perley}, {Gal-Yam}, {Groh}, {Horesh}, {Ofek}, {Kulkarni}, {Sollerman}, {Fransson}, {Rubin}, {Szabo}, {Sapir}, {Taddia}, {Cenko}, {Valenti}, {Arcavi}, {Howell}, {Kasliwal}, {Vreeswijk}, {Khazov}, {Fox}, {Cao}, {Gnat}, {Kelly}, {Nugent}, {Filippenko}, {Laher}, {Wozniak}, {Lee}, {Rebbapragada}, {Maguire}, {Sullivan}, \& {Soumagnac}}]{Yaron2017}
{Yaron}, O., {Perley}, D.~A., {Gal-Yam}, A., {et~al.} 2017, Nature Physics, 13, 510

\bibitem[{{Zhang} {et~al.}(2023){Zhang}, {Lin}, {Wang}, {Zhao}, {Li}, {Liu}, {Yan}, {Xiang}, {Wang}, \& {Bai}}]{Zhang2023}
{Zhang}, J., {Lin}, H., {Wang}, X., {et~al.} 2023, Science Bulletin, 68, 2548

\bibitem[{{Zheng} {et~al.}(2025){Zheng}, {Dessart}, {Filippenko}, {Yang}, {Brink}, {De Jaeger}, {Vasylyev}, {Van Dyk}, {Patra}, {Jacobson-Galan}, {Stewart}, {Alvarado}, {Arikatla}, {Beddow}, {Betz}, {Born}, {Bostow}, {Burgasser}, {Caceres}, {Carrasco}, {Chuang}, {DeGraw}, {Gates}, {Gendreau-Distler}, {Jacobus}, {Jennings}, {Karpoor}, {Lynam}, {Mina}, {Mora}, {Pichay}, {Ravi}, {Rees}, {Rich}, {Risin}, {Sandford}, {Savino}, {Softich}, {Theissen}, {Vidal}, {Wu}, \& {Zeng}}]{Zheng2025arXiv250313974Z}
{Zheng}, W., {Dessart}, L., {Filippenko}, A.~V., {et~al.} 2025, arXiv e-prints, arXiv:2503.13974

\bibitem[{{Zimmerman} {et~al.}(2024){Zimmerman}, {Irani}, {Chen}, {Gal-Yam}, {Schulze}, {Perley}, {Sollerman}, {Filippenko}, {Shenar}, {Yaron}, {Shahaf}, {Bruch}, {Ofek}, {De Cia}, {Brink}, {Yang}, {Vasylyev}, {Ben Ami}, {Aubert}, {Badash}, {Bloom}, {Brown}, {De}, {Dimitriadis}, {Fransson}, {Fremling}, {Hinds}, {Horesh}, {Johansson}, {Kasliwal}, {Kulkarni}, {Kushnir}, {Martin}, {Matuzewski}, {McGurk}, {Miller}, {Morag}, {Neil}, {Nugent}, {Post}, {Prusinski}, {Qin}, {Raichoor}, {Riddle}, {Rowe}, {Rusholme}, {Sfaradi}, {Sjoberg}, {Soumagnac}, {Stein}, {Strotjohann}, {Terwel}, {Wasserman}, {Wise}, {Wold}, {Yan}, \& {Zhang}}]{Zimmerman2024}
{Zimmerman}, E.~A., {Irani}, I., {Chen}, P., {et~al.} 2024, \nat, 627, 759

\end{thebibliography}

\end{document}